\documentclass[aps,twocolumn,showpacs,floats,prb, eqsecnum]{revtex4}
\usepackage[german,english]{babel}
\usepackage{amsmath,amssymb}
\usepackage{graphicx}

   \let\d=\partial

\def\nn{\nonumber}
\def\imag{i}

\let\p=\partial
\def\be{\begin{equation}}
\def\ee{\end{equation}}
\def\bea{\begin{eqnarray}}
\def\eea{\end{eqnarray}}
\def\ba{\begin{array}}
\def\ea{\end{array}}

\def\vep{{\varepsilon}}
\def\eP{{\varepsilon+P}}

\newcommand{\vq}{{\bf {q}}}
\newcommand{\vk}{{\bf {k}}}

\newcommand{\vko}{{\bf k_1}}
\newcommand{\vkt}{{\bf {k_2}}}

\begin{document}

\title{Quantum-critical relativistic magnetotransport in graphene}

\author{Markus M\"uller}
\affiliation{Department of Physics, Harvard University, Cambridge MA 02138, USA}
\author{Lars Fritz}
\affiliation{Department of Physics, Harvard University, Cambridge MA 02138, USA}
\author{Subir Sachdev}
\affiliation{Department of Physics, Harvard University, Cambridge MA 02138, USA}

\date{\today}

\begin{abstract}
We study the thermal and electric transport of a fluid of interacting Dirac fermions using a Boltzmann approach. We include Coulomb interactions, a dilute density of charged impurities and the presence of a magnetic field to describe both the static and the low frequency response as a function of temperature $T$ and chemical potential $\mu$.
In the quantum-critical regime $\mu\lesssim T$ we find pronounced deviations from Fermi liquid behavior, such as a collective cyclotron resonance with an intrinsic, collision-broadened width, and significant enhancements of the Mott and Wiedemann-Franz ratio. Some of these results have been anticipated by a relativistic hydrodynamic theory, whose precise range of validity and failure at large fields and frequencies we determine. The Boltzmann approach allows us to go beyond the hydrodynamic regime, and to quantitatively describe the deviations from magnetohydrodynamics, the crossover to disorder dominated Fermi liquid behavior at large doping and low temperatures, as well as the crossover to the ballistic regime at high fields.
Finally, we obtain the full frequency and doping dependence of the single universal conductivity $\sigma_Q$ which parametrizes the hydrodynamic response.
\end{abstract}

\maketitle


\maketitle

\section{Introduction}
Graphene attracts a lot of interest due to the massless Dirac fermions which constitute the low energy quasiparticles of the undoped material~\cite{Semenoff84,Haldane88,Zhou06,rmp}. At finite temperature and moderate doping, they form a Coulomb interacting electron-hole plasma of Dirac fermions whose transport properties are rather peculiar and significantly differ from a standard Fermi liquid.
Indeed, it has been argued that graphene is
a quantum critical system~\cite{son,guinea,ys, gorbar,joerg,FSMS},
in the sense that the inelastic scattering rate is solely set by the temperature and   proportional to $k_BT/\hbar$. To date the quantum critical and relativistic nature of such Dirac liquids, and their dynamic properties, have not been very systematically explored, despite the rather fascinating possibility it offers to study aspects of relativistic plasma physics in an easily accessible tabletop solid state system.

In a previous work~\cite{mhd}, a hydrodynamic approach describing the thermoelectric transport properties of such systems has been put forward. It predicts a highly constrained form of the frequency-dependent response functions, and their dependence on doping, magnetic field and temperature. The hydrodynamics leaves undetermined only one transport coefficient in the form of a conductivity $\sigma_Q$, whose universal value for an undoped system has recently been calculated in Ref.~\onlinecite{FSMS}. The magnetohydrodynamic analysis further suggests the presence of a collective cyclotron resonance whose origin in the quantum-critical window is a genuine many-body effect. It arises due to particles and holes colliding at a high rate with each other while collectively executing an orbiting motion at a significantly lower cyclotron frequency.

The prediction of this interesting relativistic behavior was originally initiated
by studies of a related relativistic system~\cite{nernst} which arises in the context of the superfluid-insulator transition.~\cite{bhaseen} The putative proximity of such a phase transition in the parameter space of various strongly correlated systems might significantly influence, if not dominate, their low energy physics. This
motivates the study of linearly dispersing bosonic quasiparticles which become massless at the quantum critical point. The application of Boltzmann transport theory to such critical systems was described by Damle and Sachdev~\cite{damless,ssqhe}, and has recently been extended to include the effects of a magnetic field by Bhaseen et al.~\cite{bhaseen}.
With small modifications, the analysis presented in this paper in the context of graphene can be applied to such systems, too.

The hydrodynamic analysis in Refs.~\onlinecite{nernst,mhd} relied on several assumptions, such as a weak magnetic field, light doping and weak disorder, the limits of which remained unclear. Further, the conductivity coefficient $\sigma_Q$ governing the entire frequency dependent response was only known at the quantum critical point itself, while it was not clear how the doping-driven crossover to Fermi liquid behavior~\cite{Nomura,DasSarmaGalitski} could be described. These gaps will be closed by the present analysis which, similarly to related work on graphene and other critical systems~\cite{damless,ssqhe,bhaseen,FSMS}, starts from a microscopic approach based on the Boltzmann equation, yielding an intuitive physical picture of  the crossover from quantum-critical to Fermi liquid behavior.

Not only does the present approach allow us to compute the coefficient $\sigma_Q$ and its dependence on chemical potential, frequency and magnetic field, but we will also determine the precise range of validity of the hydrodynamic analysis and the leading corrections to it. The Boltzmann approach further allows us to go beyond the hydrodynamic regime and to explore the crossover to the disorder dominated regime at large doping, where we will recover all characteristics of a Fermi liquid. We can also study the crossover out of the hydrodynamic regime to the regime of strong magnetic fields, which resolves an apparent discrepancy between hydrodynamic and Boltzmann approach concerning the value of the thermal conductivity of an undoped clean system~\cite{nernst,bhaseen}.

Several or our results on the transport in the hydrodynamic regime, on the limits of the latter and on the behavior beyond hydrodynamics turn out to be very similar to
exact results which have been obtained for strongly coupled, maximally supersymmetric conformal field theories (CFTs)~\cite{AdsCFT}. The latter can be solved exactly thanks to the AdS-CFT mapping to a weakly coupled quantum gravity problem in a universe that asymptotically becomes an Anti-de Sitter (AdS) space. It is interesting that the present weak coupling approach and the exactly soluble strongly coupled problem come to very similar conclusions regarding the limit of validity of hydrodynamics and certain aspects of the physics beyond.

\subsection{Summary of results}
We briefly highlight our main results that we hope to be experimentally observable in the near future at high enough temperatures and moderate doping where quantum interference effects and disorder are less important than the electron-electron interactions.
In the regime where $T>\mu$, graphene behaves essentially in a quantum critical manner and exhibits significant non-Fermi liquid behavior, which smoothly crosses over to the conventional Fermi liquid physics at larger doping. This is reflected in the d.c. conductivity (\ref{eq:fullsigma}) which in pure enough samples exhibits a universal, interaction-limited conductivity at low doping, and crosses over to a conductivity which grows linearly with density upon doping.
This crossover is predicted to show up in the thermopower $\alpha_{xx}$, too, with significant deviations from Mott's law in the quantum critical region, as described by Eq.~(\ref{alphatwomode}). A similar statement holds for the Wiedemann Franz law and the Lorentz ratio. We emphasize that for these effects to be seen, samples with rather low disorder are needed. Some recent experiments claim that disorder levels for graphene on substrates are currently still such that the Dirac physics is dominated by impurities~\cite{Fuhrer}. However, the recent experimental progress on suspended graphene~\cite{andrei} seems a promising route to significantly reduce disorder levels and approaching the regimes where the above non-Fermi liquid physics could be observed. The latter is expected for clean and large enough samples where the inelastic scattering dominates over impurity scattering, while the inelastic mean free path (estimated in (\ref{mfpath}), is still smaller than the sample dimensions.

Another important prediction of our paper concerns the existence of a collective cyclotron resonance  in the quantum critical regime (in all response functions), which smoothly crosses over to the standard semiclassical resonance at high doping.
Recent experiments \cite{stormer,deacon,stormerLL} have observed a ``non-hydrodynamic'' cyclotron resonance in a regime
of strong magnetic fields in which the Landau levels and their non-integer spacing in agreement with the Dirac equation can be resolved. Our prediction addresses however a very different regime at high temperatures and moderate doping where the quantization of orbits can be neglected. In this collision-dominated, semiclassical regime, the collective response of the electron-hole plasma averages over the cyclotron frequencies of non-interacting particles and holes at typical thermal energies. This translates into a resonance frequency proportional to the doped carrier density, (\ref{wc0gamma0}), occurring along with an intrinsic, interaction-mediated broadening which scales with the square of the magnetic field, cf.~Fig.~\ref{fig:cyclotronmulti}.
At high doping the resonance is predicted to turn into a sharp peak at the semiclassical value, cf. Fig.~\ref{fig:cyclotron_mu5}.

Another consequence of the magnetic field is that it renders the thermal conductivity $\kappa_{xx}$ finite, even in pure systems. At low doping there is an interesting  relationship between $\kappa_{xx}$ and the interaction-limited conductivity, which states that the thermal conductivity is inversely proportional to the zero-field conductivity, Eq.~(\ref{kappaMHDrecover}). Many of these results can be understood from a hydrodynamic point of view. However, we also predict crossovers to ballistic regimes when either the frequency or the magnetic field are increased such that the associated dynamic timescales become shorter than the inelastic scattering rate.

Our paper is structured as follows: In Sec.~\ref{model} we introduce the model of a Coulomb interacting Dirac liquid, having graphene in mind in particular. In Sec.~\ref{MHD} we briefly review the derivation of the thermo-electric response functions via a magnetohydrodynamic analysis, with emphasizing the underlying assumptions.
The formalism of the Boltzmann equation is introduced in Sec.~\ref{s:Boltzmann}. We discuss the two most relevant modes of the system, associated with charge, momentum  and energy currents, and show that due to a peculiarity of 2D systems a description restricted to these modes may give asymptotically exact results.
Sec.~\ref{s:nofield} discusses transport in the absence of magnetic fields and the dependence of the inelastic scattering rate on doping.
The hydrodynamic predictions are recovered and the leading disorder corrections are determined. The crossover to the disorder dominated Fermi liquid regime is analyzed in Sec.~\ref{s:Fermiliquid}. Finally we discuss the magnetotransport in Sec.~\ref{s:finiteB}, analyzing in detail the collective cyclotron resonance and the deviations from hydrodynamics at large fields. We conclude with a brief summary of the main results.

\section{Graphene with long-ranged Coulomb interactions and Coulomb impurities}
\label{model}
The effective low-energy description of an undoped 2 dimensional sheet of graphene is well-known to be captured by the Dirac Hamiltonian for massless electrons, where the Dirac spinor refers to the pseudospin degrees of freedom associated with the two sublattices of the carbon honeycomb lattice. In graphene, the Brillouin zone contains two inequivalent Dirac points (``valleys"), each with two spin degrees of freedom, resulting in $N=4$ species of Dirac fermions.
In this paper we consider the slightly more general situation of a liquid of weakly interacting Dirac fermions at a finite chemical potential (doping). We assume the electrons and holes to interact via standard $1/r$ Coulomb potentials, and allow for the presence of charged impurities providing long-range disorder. However, we neglect electron phonon scattering which are subdominant at low enough temperatures. Their effects are discussed, e.g., in Ref.~\onlinecite{Falkovsky07}.
The full Hamiltonian is then composed of three parts
\begin{eqnarray}
H &=& H_0 + H_1 +H_\textrm{dis},
\end{eqnarray}
where
\begin{eqnarray}
H_0 &=&- \sum_{a=1}^N\int d \mathbf{x} \left[  \Psi_a^{\dagger} \left( iv_F \vec{\sigma}\cdot \vec{\nabla}
 + \mu\right) \Psi_a \right]\;,
\end{eqnarray}
with the Fermi velocity $v_F$. The latter was measured to be approximately~\cite{Zhang05,stormer,Yacoby07} $v\approx 1.1\times 10^8 ~{\rm cm/s}\approx c/300$.
The spinor representation of the wave-function has the following Fourier decomposition
\begin{equation}
\Psi_a(\mathbf{x},t)=\int \frac{d^{2}k}{(2\pi )^{2}}\left(
\begin{array}{c}
c_{1a}(\mathbf{k},t) \\
c_{2a}(\mathbf{k},t)
\end{array}
\right) e^{i\mathbf{k}\cdot \mathbf{x}},
\end{equation}
where the operators $c_{ia}$ are the electron annihilation operators on the two different sublattices denoted $i=1,2$, and $a$ is a multi-index labeling the $N$ fermion species, i.e., spin degrees of freedom and the different valleys associated with the Dirac points.
The formulation of transport is simplest in a basis which
diagonalizes the Hamiltonian $H_{0}$. This is accomplished by a unitary transformation from the Fourier mode operators $
(c_{1a},c_{2a})$ to the basis of chiral particles $(\gamma _{+a},\gamma _{-a})$:
\begin{eqnarray}
c_{1a}(k) &=&\frac{1}{\sqrt{2}}(\gamma _{+a}(\mathbf{k})+\gamma _{-a}(
\mathbf{k})),  \notag  \label{eq:unitary} \\
c_{2a}(k) &=&\frac{K}{\sqrt{2}k}(\gamma _{+a}(\mathbf{k})-\gamma _{-a}(
\mathbf{k})).
\end{eqnarray}
We have introduced the following notation: as $\mathbf{k}$ is a two-dimensional momentum, we can define the complex number $K$ by
\begin{equation}
K\equiv k_{x}+ik_{y},~~~~\mbox{where}~~~~~\mathbf{k}\equiv (k_{x},k_{y}),
\end{equation}
and $k=|\mathbf{k}|=|K|$. Expressing the Hamiltonian $H_{0}$ in terms of $
\gamma _{\pm a}$, we obtain
\begin{equation}
H_{0}=\sum_{\lambda=\pm}\sum_ {a=1}^N\int \frac{d^{2}k}{(2\pi )^{2}}\lambda v_{F}k\, \gamma
_{\lambda a}^{\dagger }(\mathbf{k})\gamma^{\phantom{\dagger}}_{\lambda a}(\mathbf{k}).
\end{equation}
In this basis the $1/r$ interactions take the form~\cite{FSMS}
\begin{eqnarray}
&& H_1 = \sum_{a,b=1}^N \sum_{\lambda_1 \lambda_2 \lambda_3 \lambda_4} \int \frac{d^2 k_1 }{
(2 \pi )^2} \frac{d^2 k_2 }{(2 \pi )^2} \frac{d^2 q }{(2 \pi )^2}  \\
&&\times T_{\lambda_1 \lambda_2 \lambda_3 \lambda_4} (
\mathbf{k}_1 , \mathbf{k}_2 , \mathbf{q} ) \gamma_{\lambda_4 b}^{\dagger} (
\mathbf{k}_1+\mathbf{q} ) \gamma_{\lambda_3 a}^{\dagger} ( \mathbf{k}_2-
\mathbf{q} ) \nonumber \\ &&~~~~~~~~~~~~~~~~~\times \gamma_{\lambda_2 a} ( \mathbf{k}_2 ) \gamma_{\lambda_1 b} (
\mathbf{k}_1 )\,.\nn
\end{eqnarray}
Here
\begin{eqnarray}
&&T_{\lambda_1 \lambda_2 \lambda_3 \lambda_4} (\mathbf{k}_1 , \mathbf{k}_2 ,
\mathbf{q}) = \frac{V({\bf q},\omega_{\mathbf{k}_1 ,\mathbf{q}})}{8} \times  \\ &\times& \left[ 1 +
\lambda_1 \lambda_4 \frac{(K_1^{\ast} + Q^{\ast}) K_1}{|\mathbf{k}_1 +
\mathbf{q}| k_1} \right] \left[1 + \lambda_2 \lambda_3 \frac{(K_2^{\ast} -
Q^{\ast}) K_2 }{|\mathbf{k}_2 - \mathbf{q}| k_2} \right] , \nonumber  \label{deft}
\end{eqnarray}
with $\omega_{\mathbf{k}_1,\mathbf{q}}= v_F(\lambda_4|\mathbf{k}_1 +
\mathbf{q}|-\lambda_1|\mathbf{k}_1|)$, and
\bea
V({\bf q},\omega)=\frac{2\pi e^2}{\epsilon_r \epsilon(q,\omega)|{\bf q}|}
\eea
is the dynamically screened Coulomb interaction. Here $\epsilon_r$ is the dielectric constant due to the adjacent media, and $\epsilon(q,\omega)$ is the dynamic screening function which we will discuss below. Note that we have neglected the scattering between valleys $a\neq c$, since they involve large momentum transfers which are strongly suppressed.

Finally, we introduce the disorder potential
\begin{eqnarray}
H_\textrm{dis}&=& \int d \mathbf{x}V_{\textrm{dis}}(\mathbf{x})\Psi_a^{\dagger}(\mathbf{x}) \Psi^{\phantom{\dagger}}_a (\mathbf{x})\;,
\end{eqnarray}
with
\begin{eqnarray}
V_{\rm dis} ({\bf x})= \sum_{
i} \frac{Z e^2}{\varepsilon |{\bf x}-{\bf x}_i|}.
\end{eqnarray}
Here ${\bf x}_i$ denotes the random positions of charged impurities, assumed to be close to the graphene sheet, having a charge $Z e$ and average spatial density $\rho_{\rm imp}$.
Let us also express the disorder Hamiltonian $H_{\rm{dis}}$ in terms of the $
\gamma_{\lambda a}$:
\begin{eqnarray}
H_\textrm{dis}&=&\sum_{i}\sum_{a=1}^N \sum_{\lambda_1 \lambda_2} \frac{d^2 k_1 }{(2 \pi )^2} \frac{d^2 k_2 }{(2 \pi )^2} U_{\lambda_1\lambda_2} (\vko,\vkt) \\ &&\times \exp[i{\bf x}_i\cdot(\vko-\vkt)] \gamma^\dagger_{\lambda_1 a}(\vko) \gamma^{\phantom{\dagger}}_{\lambda_2 a}(\vkt),\nn
\end{eqnarray}
where
\begin{eqnarray}\label{dispot}
U_{\lambda_1\lambda_2}(\vko,\vkt)=-\frac{2\pi Z e^2}{\epsilon_r |\vko-\vkt|}\, \frac{1}{2}\left[ 1 +
\lambda_1 \lambda_2 \frac{K_1^\ast K_2}{k_1 k_2} \right]\, ,
\end{eqnarray}
which corresponds to unscreened Coulomb scatterers.
Note that even though we compute specific results for Coulomb interacting particles and Coulomb impurities, the formalism easily generalizes to arbitrary isotropic two body interactions and disorder potentials coupling to the local charge density.

\subsection{The role of screening}
It is known that in $d\leq 2$ generic interactions lead to a singularity in the amplitude for collinear forward scattering processes~\cite{chubukov,ssqhe,FSMS}. To regularize this singularity, we will  account for screening of the interactions within the random phase approximation (RPA)~\cite{gonzales,HwangSarma}. The dielectric function was calculated in Ref.~\onlinecite{HwangSarma} and has the general form
\bea
\label{epsilon}
\epsilon(q,\omega)=1+\alpha\left[k_F f_1\left(\frac{q}{k_F},\frac{\omega}{E_F}\right)+ \frac{q f_2\left(\frac{q}{k_F},\frac{\omega}{E_F}\right)}{\sqrt{1-(\frac{\omega}{v_Fq})^2}}\right],\nonumber
\eea
with $f_{1,2}$ tending to constants as $\omega,q\to 0$, and $\alpha$ is the fine structure constant of graphene:
\bea
\alpha=\frac{e^2}{\epsilon_r\hbar v_F},
\eea
which we assume to be small in the present paper, either due to a large dielectric constant, or due to its logarithmically small renormalization at low temperature.
In the static limit and at low temperatures the dielectric function reduces to
\bea
\epsilon(q,\omega=0)=1+\frac{q_{\rm TF}}{q},
\eea
where $q_{\rm TF}={\rm const. }\times\alpha k_F\sim \alpha n^{1/2}$ is the Thomas-Fermi wave vector. This leads to screening of charged impurities at finite doping, replacing the Fourier transform of the $1/r$ interactions by
$\frac{2 \pi e^2}{\epsilon_r |{\bf q}|}\to \frac{2 \pi e^2}{\epsilon_r (|{\bf q}|+q_{\rm TF})}$. However, since $q_{\rm  TF}\sim \alpha k_F$, and impurity scattering is dominated by momentum transfers of order ${\rm max}[k_F,T/v_F]$ the effect of screening is small for small $\alpha$, and will be neglected in the following. At larger $\alpha$, one would however have to deal with the screened potential of the charged impurities.~\cite{rudro,valeri}

However, screening plays a crucial role in inelastic forward scattering processes:
In the limit of collinear forward scattering where the momentum transfer tends to $\omega\to q v_F$, the second term in (\ref{epsilon}) diverges, reducing the scattering amplitude to zero. This regularizes the logarithmic divergence in the inelastic scattering crosssection. Since screening is controlled by the (small) fine structure constant $\alpha$ we will neglect it for quantitative evaluations that focus on the quantum critical regime, and only retain the screening for the purposes of regularizing the forward scattering amplitude.
Having this in mind we use the simple approximate form for the RPA screened Coulomb interaction:
\begin{eqnarray}
\label{Vscreen}
V_{\rm sc}({\bf q},\omega)= \frac{2 \pi e^2}{\epsilon_r |{\bf q}|}\frac{1}{1+\frac{\eta}{\sqrt{1-(\omega/v_Fq)^2}}},
\end{eqnarray}
where $\eta\propto \alpha$ will be taken to be an independent small parameter throughout this paper. A typical value for the proportionality constant $\eta/\alpha$ is $1/16$ which is the exact result~\cite{gonzales} for $\mu=T=0$. Having in mind small $\alpha\sim 0.1$ we will often quote numerical values calculated for a fixed value $\eta=0.01$.

We note that screening is, however, important in the Fermi liquid regime to recover standard Fermi liquid behavior. A comprehensive treatment of this regime, including dynamic screening effects, can be found in Ref.~\onlinecite{Catelani}

\section{Hydrodynamics}
\label{MHD}
We are interested in thermal and electrical transport properties of the Dirac liquid subject to interactions, disorder, as well as a perpendicular magnetic field. Each of those ingredients scatter electrons out of their linear ballistic motion, and the relative strength of these scattering processes defines various physical regimes.
We assume the external driving force (an electric field or thermal gradient) to be applied with a frequency $\omega$ which will always taken to be small compared to the largest of these scattering rates. In particular, we will not be concerned with optically driven interband transitions.

\subsection{Timescales}

The electron-electron interactions induce a finite inelastic scattering rate, which close to zero doping is of the order of
\bea
\tau^{-1}_{\rm ee}\sim {\alpha^2}\frac{k_BT}{\hbar},
\eea
and thus essentially set by the temperature. This is a hallmark of the quantum criticality of the undoped graphene system~\cite{joerg,FSMS}.
At larger doping, when the chemical potential $\mu$ exceeds $k_BT$, the inelastic scattering rate tends to the familiar Fermi liquid form $\tau^{-1}_{\rm ee}\sim T^2/\mu$, if the interactions are screened, as is the case in a Fermi liquid. As will be discussed in detail in Section~\ref{s:tauee} the scattering rate is stronger for unscreened interactions. However, we will see that only thermal transport is sensitive to the inelastic scattering rate in this Fermi liquid regime, while the electrical and mixed thermo-electrical response is dominated by other processes that are determined only by elastic scattering from impurities.

The elastic scattering rate induced by static charged impurities is naturally proportional to the density of impurities, and will be shown to be of the order of
\bea
\tau^{-1}_{\rm imp}\sim \frac{1}{\hbar}\frac{(Ze^2/\epsilon_r)^2\rho_{\rm imp}}{{\rm max}[k_B T,\mu]}.
\eea
We note that the inelastic scattering rate decreases with temperature, while the elastic scattering rate increases. The latter is due to the fact that low energy particles are more intensely scattered by Coulomb impurities.

Finally, the ``scattering rate" associated with a magnetic field is the typical cyclotron frequency of a thermally excited carrier,
\bea
\label{wctyp}
\omega^{\rm typ}_c\sim \frac{eB}{{\rm max}[k_B T,\mu]/v_F^2}\, .
\eea
Note that the cyclotron mass in the more familiar expression $\omega_c=eB/m$ is replaced by its relativistic equivalent of a typical energy divided by the square of the relevant ``speed of light" $v_F$.

\subsection{Hydrodynamic regime}
If the inelastic scattering rate $\tau^{-1}_{\rm ee}$ dominates, a hydrodynamic description of the low frequency transport should apply. This is the case at low enough doping, at high temperatures, and in moderate fields.
Indeed we will show below that the Boltzmann equation recovers precisely the predictions of relativistic magnetohydrodynamics, with the additional benefit of gaining insight into the limits of such a description. In particular, we will find that the single transport coefficient $\sigma_Q$ left undetermined by the hydrodynamic formalism, as reviewed below, is itself a function of the small parameters $\omega\tau_{\rm ee}$, $\omega^{\rm typ}_c\tau_{\rm ee}$, and $\tau_{\rm ee}/\tau_{\rm imp}$.

\subsection{Response functions from relativistic magnetohydrodynamics}
The thermoelectric response of a relativistic fluid in the presence of a magnetic field  has been derived in Refs.~\onlinecite{nernst,mhd} by a magneto-hydrodynamic analysis for the low frequency-long wavelength regime. It has been shown there~\cite{mhd} that the response at small wavevectors $k=0$ is insensitive to long-range Coulomb interactions, which is confirmed by the present analysis. Below we only briefly review the important steps and state the results for the response functions for future reference.

Hydrodynamics exploits the fact that on time scales much longer than the inelastic scattering time the only remaining dynamic modes are the diffusive currents associated with conserved quantities, i.e., charge, energy and momentum.
The linearization of the conservation laws in small deviations from equilibrium then captures the diffusive relaxation from long wavelength perturbations back to global equilibrium, and allows one to determine the frequency dependent thermoelectric response functions in the hydrodynamic regime.~\cite{km}

This program was carried out for a fluid of relativistic particles (fermions or bosons) in Refs.~\onlinecite{nernst,mhd}, and is briefly reviewed for completeness in App.~\ref{app:HD}.
The derivation of the hydrodynamic equations relies on relativistic covariance and a constitutive equation expressing the relation between electrical and thermal currents. In the limit of small thermal gradients electro-magnetic fields, the specific form of the latter is determined up to a positive coefficient $\sigma_Q$ with units of an electric conductivity, cf., (\ref{nu-constit}). Note that the validity of the  relativistic magnetohydrodynamic results is therefore restricted to small magnetic fields $B$. This will be explicitly confirmed by the more general Boltzmann theory developed in the following sections, which will result in a precise criterion for the required smallness of $B$.

The thermoelectric transport coefficients describing the current (${\bf J}$) and
heat current (${\bf Q}$) response to electric fields and temperature gradients
are defined by the relation
\begin{equation}
\left( \begin{array}{c} {\bf{J}} \\ {\bf{Q}} \end{array} \right) =
\left( \begin{array}{cc} \hat{\sigma} & \hat{\alpha} \\  T \hat{\alpha} &
\hat{\overline{\kappa}} \end{array} \right)
\left( \begin{array}{c} {\bf{E}} \\ -{\bf{\nabla}} T \end{array} \right),
\label{alltrans}
\end{equation}
where $\hat{\sigma}$, $\hat{\alpha}$ and
$\hat{\overline{\kappa}}$ are $2 \times 2$
matrices acting on the spatial $x,y$-components of the driving fields.
Rotational invariance in the plane imposes the form
\bea
\hat{\sigma}=\sigma_{xx}\, \hat{1} +\sigma_{xy}\hat{\epsilon},
\eea
where $\hat{1}$ is the identity, and $\hat{\epsilon}$ is the antisymmetric
tensor with $\hat{\epsilon}_{xy}=-\hat{\epsilon}_{yx}=1$.
Note that the two off-diagonal entries in (\ref{alltrans}) are related due to Onsager reciprocity.

The thermal conductivity, $\hat{\kappa}$, defined as the heat
current response to $-\vec{\nabla} T$ in the absence of an electric
current (electrically isolated boundaries), is given by
\begin{equation}
\hat{\kappa} = \hat{\overline{\kappa}} -T \hat{\alpha} \hat{\sigma}^{-1}
\hat{\alpha}. \label{kappadef}
\end{equation}

The analysis of the hydrodynamic equations in a weak magnetic field yields the following frequency dependent response.~\cite{mhd,nernst} The longitudinal and Hall conductivity are
\bea
\sigma_{xx}(\omega)&=& {\sigma }_Q \frac{\omega
\left( \omega + i\gamma  + i \omega_c^2/\gamma  \right) }{
\left(\omega + i\gamma \right)^2 -\omega_c^2  }\,, \label{sxxf} \\
\sigma_{xy}(\omega)&=& -\frac{\rho}{B}
\frac{\omega_c^2+\gamma^2-2i\gamma\omega}
{ \left(\omega + i\gamma \right)^2 -\omega_c^2}. \label{sxyf}
\end{eqnarray}
The thermopower and transverse Peltier coefficient are found to be
\begin{eqnarray}
\alpha_{xx}(\omega) &=& -\frac{\omega\left[ \sigma_Q
(\mu/T)(\omega+i\gamma) -i s  \rho/(\varepsilon+P)\right]}{ \left(\omega + i\gamma \right)^2 -\omega_c^2  }\,,\quad\\
\alpha_{xy}(\omega) &=& -\frac{s}{B}
  \frac{\omega_c^2+\gamma^2 -i\gamma \omega [1-
\mu\rho/(s T)]}{ \left(\omega + i\gamma \right)^2 -\omega_c^2},
\eea
and the thermal conductivities
\bea
\overline{\kappa}_{xx}(\omega) &=&
\frac{-\gamma \frac{\eP}{T} -
    i \frac{s^2 T}{\eP}\, \omega    +
     \sigma_Q \frac{\mu^2}{T} \omega
     \left(\omega+ i\gamma\right) }{\left( \omega+
i\gamma\right)^2 -\omega_c^2}\,,\nn \\
\overline{\kappa}_{xy}(\omega) &=&
-\frac{B}{T}\frac{\frac{s^2\,T^2\,\rho  }{{\left(\eP  \right) }^2}
- \mu \sigma_Q \,\left[ \gamma\frac{\mu\rho}{\eP}   - 2i\frac{s\,T}{P
+ \varepsilon }
 \left( \omega + i\gamma\right) \right]}{
{\left(\omega + \imag \,\gamma\right) }^2 -
    {{{\omega }_c}}^2}\,,\nn\\
\label{kappaMHD}
\kappa_{xx}(\omega) &=&  i\frac{(\varepsilon+P)}{T} \frac{(\omega
+i\omega_c^2/\gamma)}{(\omega  + i
\omega_c^2 /\gamma)^2 - \omega_c^2}, \\
\kappa_{xy}(\omega) &=& \frac{(\varepsilon+P)}{T} \frac{\omega_c}{(\omega + i \omega_c^2/\gamma)
^2 - \omega_c^2}.
\eea
In these formulae a collective cyclotron frequency $\omega_c$ and a damping rate $\gamma$ have been defined as
\begin{equation}
\label{omegac}
\omega_c \equiv \frac{e B \rho v_F^2}{(\varepsilon + P)}~~~;~~~\gamma \equiv
\frac{\sigma_Q B^2 v_F^2}{(\varepsilon + P)}\,,
\end{equation}
where here $B$ is given in SI units. To obtain results in more customary cgs units one should replace $B\to B/c$ throughout the paper. This shows that the speed of light merely plays the role of a coupling constant determining the strength of the magnetic field.
By introducing a phenomenological relaxation rate $\tau^{-1}_{\rm imp}$ (due to weak impurity scattering) into the momentum conservation law, one finds that the above formulae are simply changed by the replacement $\omega\to \omega+i/\tau_{\rm imp}$.
However, since disorder breaks explicitly the relativistic invariance by singling out its own rest frame, it is to be expected that the resulting expressions for the response are merely qualitatively correct and become exact only in the limit when the elastic scattering rate is the smallest rate in the problem. This will indeed be confirmed below.

It will be one of our aims in the subsequent sections to rederive these response
functions in the appropriate hydrodynamic regime, and to establish the precise limits of validity of the hydrodynamic description, and in particular the admissible range of $B$ and $\omega$. Further we will obtain explicit expressions for the transport coefficient $\sigma_Q$ as a function of the system parameters, most importantly as a function of the chemical potential $\mu$.
This will shed light on the crossover from relativistic, quantum critical response in the regime $|\mu|\lesssim T$ where both particles and holes contribute to transport - to the Fermi liquid regime $|\mu|\gg T$ where only one kind of particles contributes. In the latter regime we will recover the standard laws governing Fermi liquids.
 We will also study the disorder dependence of the response functions and discuss the crossover from the interaction dominated to the disorder dominated regime.

\section{Boltzmann transport}
\label{s:Boltzmann}

\subsection{Applicability of Boltzmann transport theory}

For the Boltzmann equation to be valid, one requires the existence of well-defined, sharp quasiparticle excitations and sufficiently weak interactions so that scattering does not lead to strong many body correlations. This framework is of course much more general than hydrodynamics. We therefore can extend our study of transport into regimes of strong magnetic fields and disorder.
However, we are always restricted to a regime where $k_BT$ is much bigger than the cyclotron energy of thermal particles, {\it{i.e.}}, $k_B T \gg \hbar \omega_c^{\rm typ}$. This ensures that we do not need to account for Landau quantization of electron orbits and the quantum Hall effect, which lies beyond the semiclassical Boltzmann equation. Similarly, localization corrections which derive from quantum interference cannot be captured by a simple Boltzmann approach.
However, as long as the interference effects do not drive the system insulating, as ascertained in several recent studies of non-interacting electrons, at least for random point-like disorder~\cite{nonlocRyu,nonlocBrouwer}, one can expect that the Boltzmann approach applied to a field theory with appropriately renormalized parameters for interactions and disorder~\cite{herbut,FosterAleiner} will capture a large part of the phase diagram in temperature, disorder and interaction strength. Such an approach was recently taken to predict a logarithmic increase with temperature of the conductivity in clean undoped graphene~\cite{FSMS}.

The central object in Boltzmann transport theory is the distribution matrix of the quasiparticles
\begin{equation}
f_{\lambda \lambda'}(\mathbf{k},t)=\left\langle \gamma _{\lambda a}^{\dagger }(
\mathbf{k},t)\gamma^{\phantom{\dagger}}_{\lambda' a}(\mathbf{k},t)\right\rangle \; ,  \label{defg}
\end{equation}
where there is no sum over $a$ on the RHS, and we assume the distribution
functions to be the same for all valleys and spins. For all further discussions we will neglect the matrix elements off-diagonal with respect to the helicity basis labeled by $\lambda$, which brings us back to the familiar quasiparticle distribution function
\begin{equation}
f_{\lambda }(\mathbf{k},t)=\left\langle \gamma _{\lambda a}^{\dagger }(
\mathbf{k},t)\gamma^{\phantom{\dagger}}_{\lambda a}(\mathbf{k},t)\right\rangle .
\end{equation}
This approximation can safely be made since we are interested in low frequencies $\hbar \omega\ll k_BT$ where field-induced, coherent interband transitions leading to off-diagonal correlations (with $f_{\lambda\neq\lambda'}$) can be neglected.
In equilibrium, {\em i.e.\/}, in the
absence of external perturbations, the distribution functions are Fermi functions
\begin{eqnarray}
f_{\lambda}(\mathbf{k},t) &=&f^{0}_\lambda (k)= \frac{1}{e^{(\lambda v_F k -\mu)/T}+1},
\end{eqnarray}
at the finite chemical potential $\mu$, as defined by the doping or gate potential.

We consider the Boltzmann equation in the presence of an electrical field ${\bf{E}}$, the Lorentz force due to a perpendicular magnetic field ${\bf{B}}=B{\bf e}_z$, and a spatially varying temperature $T({\bf{r}})$:
\bea
&&\left( \frac{\partial }{\partial t}+e \left( \mathbf{E}+ {\bf{v}}_{\lambda,{\bf{k}}} \times \mathbf{B}\right)\cdot \frac{\partial }{
\partial \mathbf{k}} + {\bf{v}}_{\lambda,{\bf{k}}}\cdot \frac{\partial }{
\partial \mathbf{r}}\right) f_{\lambda }(\mathbf{r},\mathbf{k},t)\nn\\
&&\quad\quad\quad\quad\quad\quad\quad\quad\quad=-{\cal I}_{\rm coll}[\lambda,\mathbf{r},\mathbf{k},t\,|\{f\}].  \label{trans0}
\eea
Here ${\cal I}_{\rm coll}[\lambda,\mathbf{r},\mathbf{k},t\,|\{f\}]$ denotes the collision integral due to Coulomb interactions and impurity scattering, and ${\bf{v}}_{\lambda,{\bf{k}}}=\nabla_{\bf k}\varepsilon_{\lambda\bf k}=\lambda v_F{\bf e_k}$ with ${\bf e_k}=({\bf k}/k)$, denotes the quasiparticle velocity.
We rewrite the equation specifying the driving gradients on the RHS:

\begin{eqnarray}
&&\partial_t f_{\lambda }(\mathbf{r},\mathbf{k},t) + e \lambda v_F B \left({\bf{e}}_{\bf{k}}\times {\bf{e}}_z\right)\cdot \frac{\partial}{\partial {\bf{k}}} f_{\lambda }(\mathbf{r},\mathbf{k},t)\nn\\
&&\quad +{\cal I}_{\rm coll}[\lambda,\mathbf{r},\mathbf{k},t\,|\{f\}] =\mathcal{{\bf{F}}}^{E}\cdot e {\bf{E}}+\mathcal{{\bf{F}}}^T \cdot{\bf {\nabla}}T,
\end{eqnarray}

where
\begin{eqnarray}
\mathcal{{\bf{F}}}^E&=& \lambda \frac{v_F}{T} {\bf{e}}_{\bf{k}} f^0_\lambda(k)[1-f^0_\lambda(k)],
\eea
and
\bea
\mathcal{{\bf{F}}}^T&=& -\frac{v_F(v_F k-\lambda \mu)}{T^2} {\bf{e}}_{\bf{k}} f^0_\lambda(k)[1-f^0_\lambda(k)] \;.
\end{eqnarray}
We seek to solve (\ref{trans0}) in linear response and thus
parameterize the deviation of $f_{\lambda }({\bf r},{\bf k},t)$ from its equilibrium value in the standard way~\cite{ziman} as
\begin{widetext}
\bea
f_{\lambda }(\mathbf{r},\mathbf{k},\omega )&=&2\pi \delta (\omega )f^{0}_\lambda(
k,T(\mathbf{r}))  + f^{0}_{\lambda k}[1-f^{0}_{\lambda k}] \frac{v_F}{T^2}  \mathbf{e_k}\cdot\left[ e\mathbf{E}(\omega ) g^{(E)}_{\parallel,\lambda}\left (\frac{v_Fk}{T},\omega \right) + \nabla T(\omega ) g^{(T)}_{\parallel,\lambda}\left(\frac{v_Fk}{T},\omega \right)\right]  \nn\\
&+& f^{0}_{\lambda k}[1-f^{0}_{\lambda k}] \frac{v_F}{T^2}  \left({\bf{e}}_{\bf{k}}\times\mathbf{e}_z\right) \cdot\left[ \mathbf{E}(\omega ) g^{(E)}_{\perp,\lambda}\left (\frac{v_Fk}{T},\omega \right) + \nabla T(\omega )  g^{(T)}_{\perp,\lambda}\left (\frac{v_Fk}{T},\omega \right)\right],
\eea
\end{widetext}
with $T(\mathbf{r})=T+{\bf{r}}\cdot \nabla T(\omega)$ and dimensionless functions $g_\lambda(k)$.

We note that exactly at particle hole symmetry ($\mu=0$), an applied electric field
generates perturbations $g_\lambda^{(E)}$ having opposite sign for
quasiparticles and quasiholes,
\bea
\label{phsym}
g^{(E)}_\lambda\left (\frac{v_Fk}{T},\omega \right )=\lambda g^{(E)}\left (\frac{v_Fk}{T},\omega \right),
\eea
whereas a thermal gradient will generate symmetric perturbations,
\bea
\label{phsymT}
g^{(T)}_\lambda \left (\frac{v_Fk}{T},\omega \right )= g^{(T)}\left (\frac{v_Fk}{T},\omega \right )\;.
\eea
However, in the case of a finite chemical potential the distribution function will have a generic dependence on $\lambda$. Notice also that the perpendicular components of the perturbations, $g_{\perp,\lambda}$, vanish in the absence of  a magnetic field.

\subsection{Matrix formalism}
\label{s:matrix}
In this section we will set up the calculational framework for all subsequent discussions. A standard way to deal with integro-differential equations consists in expanding the solution $g_\lambda$ into a set of basis functions $\phi_n(\lambda,k)$,
\begin{eqnarray}
\label{basisexpansion}
g_{||,\lambda}(k)&=&\sum_{n} \psi_n^{||} \phi_n(\lambda,k) \; ,\nonumber \\ g_{\perp,\lambda}(k)&=&\sum_{n} \psi_n^{\perp} \phi_n(\lambda,k) \; ,
\end{eqnarray}
and to express the integral equation as a matrix equation, by multiplying it from the left with different basis functions and integrating and summing over $k$ and $\lambda$, respectively.

From now on we take momenta $k$ to be given in units of $\frac{v_F}{T}$ unless stated otherwise. More generally, we will use units in which $\hbar =k_BT=v_F=1$, but restore those in final results.
The scattering terms then turn into a matrix acting in the space of expansion coefficients $\psi_n^{||}$ and $\psi_n^\perp$, which we organize into a doublet of vectors
\begin{eqnarray}
\vec{\psi}=\left ( \begin{array} {c}  \vec{\psi}^{||} \\ \vec{\psi}^\perp \end{array}\right)\; ,
\end{eqnarray}
allowing us to cast the Boltzmann equation into the compact form:
\begin{eqnarray}\label{matrixboltz}
\left ( \begin{array}{cc} \mathcal{M} & -\mathcal{B}\\ \mathcal{B} & \mathcal{M }\end{array}\right)\cdot \vec{\psi}^{E,T} = \left ( \begin{array}{c} \vec{\mathcal{F}}^{E,T}\\0\end{array} \right) \;.
\end{eqnarray}
We have used that in linear response we can deal with the response to the electric field and the thermal gradient separately.
A matrix inversion yields the solution
\begin{eqnarray}\label{eq:matrix}
\vec{\psi}^{E,T}=\left(\begin{array}{cc}  K & \overline{K} \\ -\overline{K}  & K \end{array} \right)\cdot \left ( \begin{array}{c} \vec{\mathcal{F}}^{E,T}\\0\end{array} \right) \; ,
\end{eqnarray}
with
\begin{eqnarray}
K&=&\left(\mathcal{M}+\mathcal{B}\mathcal{M}^{-1}\mathcal{B}\right)^{-1} \; ,\nonumber \\ \overline{K}&=&\mathcal{M}^{-1} \mathcal{B}\left(\mathcal{M}+\mathcal{B}\mathcal{M}^{-1}\mathcal{B}\right)^{-1} \;.
\end{eqnarray}
In the above, the matrix $\mathcal{M}$ is the sum of three terms:
\bea
\mathcal{M}=\mathcal{M}^{\rm{Cb}}+\mathcal{M}^{\rm{imp}}+\mathcal{M}^{i\omega},
\eea
the first accounting for inelastic scattering due to Coulomb interactions, the second describing impurity scattering, while the last derives from the time derivative in the Boltzmann equation.

Let us consider the first two terms, corresponding to the collision term on the right hand side of (\ref{trans0}). For sufficiently weak interactions and for dilute enough impurities, the collision integral is given by an application of Fermi's golden rule to two body collisions, and to the scattering from Coulomb impurities~\cite{ssqhe}. The corresponding matrix elements are given in App.~\ref{app:Melements}.
The strength of the inelastic scattering rate due to electron-electron interactions is characterized by
$\alpha^2$ while elastic scattering rate from impurities is measured by the dimensionless parameter
\begin{eqnarray}
\Delta=\pi^2 \left ( \frac{Z e^2}{k_BT\epsilon_r}\right)^2 n_{{\rm {imp}}}\,.
\end{eqnarray}

The magnetic field deflects particles from linear propagation at a rate proportional to
 the dimensionless parameter characterizing the magnetic field strength
\begin{eqnarray}
b=\frac{eBv_F^2}{(k_BT)^2}\;.
\end{eqnarray}

The relative magnitude of $\alpha^2$, $\Delta$ and $b$ defines various transport regimes which we will discuss below.

\subsection{Linear response}
The heat current (${\bf{Q}}$) is related to the energy current (${\bf {J}}^E$) and the electrical current (${\bf{J}}$) via
\begin{eqnarray}
{\bf{Q}}={\bf{J}}^E-\frac{\mu}{e} {\bf{J}} \; .
\end{eqnarray}
Given a perturbation of the distribution function parametrized by $\vec{\psi}_{\parallel,\perp}$, the associated heat and electrical currents are given by the expressions
\bea
\label{currents1}
\left(\begin{array}{c} \mathbf{J}_{\parallel,\perp} \\ \mathbf{Q}_{\parallel,\perp}
\end{array} \right)
&=& N \sum_\lambda\int \frac{d^2k}{(2\pi)^2} \frac{k_x^2}{k^2} f^0_{\lambda k}(1-f^0_{\lambda k})\\
&& \quad \times\sum_m \psi_{\parallel,\perp;m} \phi_m(\lambda,k)
\left( \begin{array}{c} e \lambda \\ k-\lambda \mu
\end{array} \right),\nn
\eea
where the shorthand
\begin{eqnarray}
f^0_{\lambda k}:=\frac{1}{e^{\lambda k-\mu}+1}
\end{eqnarray}
denotes the Fermi distribution.

Using the matrix elements of the driving terms, cf., (\ref{eq:force}), we can express this as
\bea
\label{currents}
\left(\begin{array}{c} \mathbf{J}_{\parallel,\perp} \\ \mathbf{Q}_{\parallel,\perp}
\end{array} \right)
&=& \frac{N}{2}\sum_m \psi_{\parallel,\perp;m}
\left( \begin{array}{c} e{\cal F}_m^E \\ -{\cal F}_m^T
\end{array} \right),
\eea

Using the formal solution (\ref{eq:matrix}) for the coefficients $\psi$,
we immediately read off the longitudinal transport coefficients defined in (\ref{alltrans}):
\begin{eqnarray}\label{eq:coeff}
\sigma_{xx}(\omega)&=&\frac{J_x(\omega)}{E_x(\omega)}=\frac{N e^2}{2 \hbar} \vec{\mathcal{F}}_E  \cdot K \vec{\mathcal{F}}_E,   \nonumber \\ \alpha_{xx}(\omega)&=&-\frac{J_x(\omega)}{\nabla_x T(\omega)}=-\frac{Ne k_B}{2\hbar}  \vec{\mathcal{F}}_E \cdot K \vec{\mathcal{F}}_T  , \nonumber \\  \overline{\kappa}_{xx}(\omega)&=& \frac{Q_x (\omega)}{-\nabla_x T(\omega)}=\frac{Nk_B^2 T}{2\hbar}  \vec{\mathcal{F}}_T \cdot K \vec{\mathcal{F}}_T.
\label{resppara}
\end{eqnarray}
In the presence of a magnetic field, the transverse transport coefficients are finite as well and given by
\begin{eqnarray}
\sigma_{xy}(\omega)&=&\frac{J_x(\omega)}{E_y(\omega)}=\frac{N e^2}{2 \hbar} \vec{\mathcal{F}}_E  \cdot \overline{K} \vec{\mathcal{F}}_E \; , \nonumber \\ \alpha_{xy}(\omega)&=&-\frac{J_x(\omega)}{\nabla_y T(\omega)} = -\frac{N e k_B}{2 \hbar} \vec{\mathcal{F}}_E \cdot  \overline{K} \vec{\mathcal{F}}_T   \; ,\nonumber \\ \overline{\kappa}_{xy}(\omega)&=&\frac{Q_x (\omega)}{-\nabla_y T(\omega)}=\frac{N k_B^2 T}{2 \hbar} \vec{\mathcal{F}}_T \cdot \overline{K} \vec{\mathcal{F}}_T  \; .
\label{respperp}
\end{eqnarray}

\subsection{Choice of basis}
In order to analyze the response functions, it proves essential to choose a well-adapted basis $\phi_{n=0,\dots,{\cal N}}(\lambda,k)$ to expand $g_{\parallel,\perp;\lambda}(k)$ into.

The structure of the currents and of  the driving terms (\ref{currents}) suggests to use the modes
\bea
\label{modes}
\phi_0(\lambda,k)=k,\\
\phi_1(\lambda,k)=\lambda.
\eea
Moreover, it will be convenient to complete the basis in such a way
that the $\phi_{n\geq 2}$ do not contribute to the electrical and thermal currents.
Due to reciprocity, this implies in turn, that these modes do not couple to the driving fields, i.e., ${\cal F}^{E,T}_{n\geq 2}=0$, or
\begin{eqnarray}
\label{orthogonality}
\vec{{\cal F}}^{E,T}= \left ( \begin{array}{c} \vec{{\cal F}}^{E,T}_0 \\ \vec{{\cal F}}^{E,T}_1 \\ \vec{0} \end{array}\right).
\end{eqnarray}

According to (\ref{eq:force}) this basis choice imposes the following two constraints for all $n\geq 2$
\bea
\label{ortho}
\sum_\lambda\int d^2k f^0_{\lambda k}(1-f^0_{\lambda k}) \phi_{n\geq 2}(\lambda,k)\phi_{0,1}(\lambda,k)=0.
\eea
From the expressions in App.~\ref{app:Melements} it is easy to check that this implies the vanishing of the matrix elements
\bea
\label{M0n}
{\cal M}^{i\omega}_{0,n}&=&{\cal M}^{i\omega}_{1,n}=0,\nn\\
{\cal B}_{0,n}&=&0,\quad {\textrm {for all}}\; \quad n\geq 2.
\eea
This will play a vital role in establishing the relativistic magnetohydrodynamics in Sec.~\ref{s:finiteB}.

Note that the part of the collision integral due to elastic scattering from disorder does not follow the same pattern. Rather it has non-vanishing matrix elements ${\cal M}^{\rm imp}_{0n}$ for all $n$. This is closely related to the breaking of relativistic invariance by the disorder.

The mode $\phi_0(\lambda,k)$ is central not only because it naturally describes the energy current, but it also has the further important property that it is the deviation generated by a transformation to a moving frame, and translational invariance protects this mode from decaying due to Coulomb interactions. Indeed, one immediately checks that momentum conservation implies that $\phi_0$ is a zero mode of the Coulomb collision operator, see Eq.~(\ref{MCb}),
\bea
{\cal M}^{\rm Cb}_{0n}={\cal M}^{\rm Cb}_{n0}=0\quad  {\textrm{for all}} \quad n.
\eea
However, the momentum is not conserved by impurity scattering and the magnetic field.

The relativistic linear dispersion is essential to ensure that the ``momentum mode" $\phi_0$ coincides with the ``energy mode" entering the energy and heat currents. Moreover, it is because of this relativistic dispersion that the scattering due to the magnetic field couples $\phi_0$ only to the ``electrical current mode" $\phi_1$. This structure in the Boltzmann equation is crucial to retrieve the relativistic magnetohydrodynamics, as will become clear in Sec.~\ref{s:finiteB}.

\subsection{Leading logarithmic approximation}
\label{sec:leadlog}
A significant simplification occurs in two dimensions, as noticed
in a preceding publication~\cite{FSMS}. Indeed, a close analysis of the two body collision integral (\ref{MCb}) shows that there is a logarithmic divergence in the phase space for nearly collinear forward scattering processes~\cite{ssqhe,FSMS} which is only cut off at small angles of order ${\cal O}(\alpha)$, e.g., by the RPA screening of the Coulomb interactions, see Eq.~\eqref{Vscreen}. Interaction corrections to the linear dispersion or the inclusion of a finite life time of the quasiparticles would provide a similar cut-off, too.
In the limit of weak interactions $\alpha\ll 1$, but still in the hydrodynamic regime (assuming even smaller impurity scattering rates, frequencies and magnetic fields, $\Delta,\omega/T,b\ll \alpha^2$), this leads to an equilibration among excitations moving in the same spatial direction. The off-equilibrium distribution function will then be characterized by an angle-dependent effective temperature and chemical potential. Since the angular dependence in linear response has to be proportional to the projection of $\mathbf{k}$ onto the driving field, such perturbations to the distribution functions correspond precisely to the two modes $\phi_0$ and $\phi_1$ introduced in the previous section~\cite{FSMS,kashuba}. Provided the logarithmic anomaly in the forward scattering is sufficiently strong the above justifies to restrict the analysis of the Boltzmann equation to these two modes.

To be more precise, we actually invoke that the inelastic relaxation of all other modes $\phi_{n\geq 2}$ is faster by a factor of the order of $\log(1/\eta)$ than that of the "electrical current mode" $\phi_1$, where $\eta\propto \alpha$ is the screening parameter introduced in (\ref{Vscreen}).
Up to logarithmically small admixtures of other modes, $\phi_1$ corresponds indeed to the eigenvector of ${{\cal M}}^{\rm Cb}$ with the smallest positive eigenvalue.

It follows that we may obtain exact solutions of the Boltzmann equation to leading order in $[\log(1/\eta)]^{-1}$ by restricting ourselves to the Ansatz
\begin{eqnarray}
\label{ansatzgh}
g_\lambda (k,\omega)&=& \psi_0 (\omega)\phi_0(\lambda,k) + \psi_1(\omega)\phi_1(\lambda,k)\nonumber \\ &=& \psi_0 (\omega)k +  \psi_1(\omega)\lambda,
\end{eqnarray}
 for the deviation of the distribution function.
This is demonstrated in Fig.~\ref{fig:sigmaofeta} where we plot $\alpha^2\sigma(\mu=0)$ as a function of $\eta$. The lower data set was evaluated within the two mode approximation as in Ref.~\onlinecite{FSMS}, while the upper curve was obtained by solving the Boltzmann equation projecting onto 12 basis functions $\phi_n(\lambda,k)$ and inverting the resulting matrices.~\cite{footnote1}
It is clearly seen that in the limit of weak screening $\eta\ll 1$, the two mode approximation becomes exact.
\begin{figure}
\includegraphics[width=0.45\textwidth]{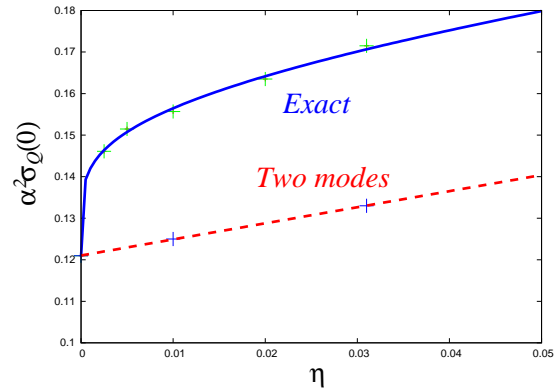}
\caption{Dependence on the screening parameter $\eta$ of the conductivity of undoped graphene, $\alpha^2\sigma(\mu=0;\eta)$ in units of $e^2/\hbar$. At very small $\eta$, $\alpha^2\sigma(0;\eta)$ approaches its limiting value~\cite{FSMS} $0.121$ as $\alpha^2[\sigma(0;\eta)-\sigma(0;0)]\sim [\log(1/\eta)]^{-1}$, the solid line being a fit to $a+b[\log(1/\eta)]^{-1}$. The lower curve is a linear fit to the data obtained from the two mode approximation, which becomes asymptotically exact as $\eta\to 0$.}
\label{fig:sigmaofeta}
\end{figure}

This two mode approximation is well justified in the hydrodynamic regime where the inelastic scattering dominates the dynamics. However, we will argue in Section~\ref{sec:tma} that the Ansatz (\ref{ansatzgh}) also
captures the relaxation time approximation which is widely used in the disorder dominated Fermi liquid regime $\mu\gg T$. In the case where the inelastic scattering can be neglected, the relaxation time approximation even becomes exact, boiling down to using the single mode $\phi_1$. We will thus use the Ansatz (\ref{ansatzgh}) later to describe analytically the crossover from interaction-dominated to disorder-dominated regime. However, for most explicit results, and for the entire analytical discussion of magnetotransport we will not resort to this approximation. Nevertheless, it will become clear that the essential physics is captured by the dynamics of the two modes $\phi_0,\phi_1$ which carry the information of energy and charge currents, respectively. The matrix elements pertaining to these two modes are given explicitly in App.~\ref{app:Melements} and will be used in the discussion of transport in the remaining sections.

\section{Transport in the absence of a field}
\label{s:nofield}
As we will see in later sections, it is essentially the slow dynamics of the momentum mode $\phi_0$ which gives rise to the special relativistic structure of the response functions in the hydrodynamic regime.
It is thus convenient to treat the zero mode $\phi_0$ separately, and write the matrix ${\cal M}={\cal M}^{\rm Cb}+{\cal M}^{i\omega}+{\cal M}^{\rm imp}$ in the form
\bea
{\cal M}=\left(\begin{array}{ccccc}
M_0&M_1&M_2&\cdots&M_n\\
M_1&\mathbf{M}_{11}&\mathbf{M}_{12}&\cdots&\mathbf{M}_{1n}\\
M_2&\mathbf{M}_{21}&\mathbf{M}_{22}&\cdots&\mathbf{M}_{2n}\\
\vdots&\vdots& & &\vdots\\
M_n&\mathbf{M}_{n1}&\mathbf{M}_{n2}&\cdots&\mathbf{M}_{nn}
\end{array} \right)\,.
\eea
Owing to (\ref{orthogonality}), we only need to know the inverse of this matrix in the sector spanned by $\phi_{0},\phi_1$ in order to calculate the response functions (\ref{resppara},\ref{respperp}). In this sector it assumes the form
\bea
{\cal M}^{-1}&=&\frac{1}{M_0-\sum_{n\geq 1}M_nM_m (\mathbf{M}^{-1})_{nm}}\left(\begin{array}{ccccc}
1&a&*&\cdots&*\\
a& c&*&\cdots&*\\
*&*&*&\cdots&*\\
\vdots&\vdots& & &\vdots\\
*&*&*&\cdots&*
\end{array} \right)\,, \nn
\eea
with
\bea
a&=&-\sum_{n\geq 1}M_n(\mathbf{M}^{-1})_{1n}\\
c&=&M_0(\mathbf{M}^{-1})_{11}-\sum_{m,n\geq 2}\left(
M_nM_m\frac{\det[\mathbf{M}^{(1n,1m)}]}{\det \mathbf{M}}\right),\nn
\eea
where $\det[\mathbf{M}^{(1n,1m)}]$ are the subdeterminants of the matrix $\mathbf{M}$ where the two rows $1$ and $n$, and the two columns $1$ and $m$ have been dropped. The above expressions are very useful in the hydrodynamic limit, where typical matrix elements of ${\bf M}$ are much larger than the entries $M_n$.

\subsection{Clean case}
\label{s:tauee}
We recall that in the absence of disorder all $M_{n\geq 2}$ vanish, which simplifies the above expression for the matrix inverse to
\bea
\label{cleanMinverse}
{\cal M}^{-1}&=&\frac{1}{M_0-M_1^2 g_1}\left(\begin{array}{ccccc}
1&-M_1g_1&*&\cdots&*\\
-M_1g_1& M_0 g_1&*&\cdots&*\\
*&*&*&\cdots&*\\
\vdots&\vdots& & &\vdots\\
*&*&*&\cdots&*
\end{array} \right)\,, \nn\\
g_1&\equiv &g_1(\mu,\omega)=(\mathbf{\mathbf{M}}^{-1})_{11} \equiv \frac{\hat{g}_1}{\alpha^2}\; .
\eea

Note that in this case the judicious choice of the basis $\phi_n$ allows us to summarize the effect of all the modes $\phi_{n\geq 2}$ into a single (frequency dependent) matrix element $g_1$, which enters all the response functions. $g_1$ characterizes the inelastic scattering rate due to electron-electron interactions as we will detail below. In the case where the inelastic scattering rate dominates, it is convenient to write $g_1\equiv \hat g_1/\alpha^2$ to exhibit explicitly the scaling with interaction strength, $\hat g_1$ being a number of order ${\cal O}(1)$.

In general it is sufficient to use a relatively small number of bases to obtain a good accuracy for $g_1$.
However, as we discussed in Section~\ref{sec:leadlog}, in the hydrodynamic regime in two dimensions, there is even a further simplification that allows us to neglect the remaining modes $\phi_{n\geq 2}$ to leading order in $\log(1/\eta)$.

From Eqs.~\eqref{eq:matrix},~\eqref{eq:force}, and~\eqref{eq:coeff} it is easy to show that the response in a clean system is given by
\begin{eqnarray}
\label{sigmapure}
&&\sigma_{xx}(\omega;\mu,\Delta=0)=  e^2\frac{\rho^2v_F^2}{\eP}\frac{1}{(-i\omega)}+\sigma_Q,
\end{eqnarray}
where $\sigma_Q=\sigma_Q(\mu,\omega)$ is the $\omega$- and $\mu$-dependent  transport coefficient
\bea
\label{sigmaQmu}
\sigma_Q(\mu,\omega)=\frac{e^2}{\hbar}\frac{1}{\alpha^2}\frac{N}{2 \hat g_1}\left(\tau_{\rm ee}\frac{\alpha^2k_B T}{\hbar}\right)^2
 \frac{1}{1-i\omega\tau_{ee}},
\eea
which was left undetermined in the hydrodynamic formalism, cf.~(\ref{sxxf}).
In the above we have defined the inelastic scattering rate as
\bea
\label{tauee}
\tau_{\rm ee}^{-1}=\alpha^2\frac{N}{2{\hat g_1}}\frac{k_B T}{\hbar} \left[ \frac{N\ln[2\cosh(\mu/2)]}{2\pi}-\frac{\rho^2(\hbar v)^2}{(\eP)T}\right]^{-1}\,.
\eea

\subsection{Inelastic scattering rate in the Fermi liquid regime $\mu\gg T$}
In the above expressions, $\hat g_1$ is a scaling function of $\mu/T$ and $\omega \tau_{\rm ee}$. As long as we concentrate on the hydrodynamic frequency regime, the latter is negligible. However, the $\mu$-dependence of $\hat g_1$ and $\tau_{\rm ee}$ are important.

An order of magnitude estimate for $g_1$ can be obtained from the inverse of the expression (\ref{W}) for $\mathbf{M}^{\rm Cb}_{11}$. If the electron-electron interactions are not screened (as is the case to lowest order in $\alpha$), this multiple integral saturates to a finite value at large $\mu/T$. This is explained in detail in App.~\ref{app:inel}, and is borne out by direct numerical evaluation,
\bea
\label{g_1estimate}
g_1^{-1}(\mu\gg T)= {\cal O}(1).
\eea
The inelastic scattering rate for unscreened long range interactions can similarly be estimated to scale as
\bea
\tau^{-1}_{\rm ee} \sim \alpha^2\mu,
\eea
as follows from the above together with (\ref{tauee}).

If we include screening of the interactions one finds instead
\bea
\label{g_1estimate_screen}
g_{1,\rm sc}^{-1}(\mu\gg T)= {\cal O}\left(\alpha^2\frac{T^2}{\mu^2}\right),
\eea
and the analogous estimate for $\tau_{\rm ee}^{-1}$ yields
the familiar Fermi liquid behavior
\bea
\label{taueesc}
\tau^{-1}_{\rm ee,sc}\sim \alpha^2\frac{T^2}{\mu}\,.
\eea

We have calculated the full function $g_1^{-1}(\mu)$ in the limit of vanishing screening (i.e., $\eta\to 0$), where the evaluation via the formula for the matrix element~\eqref{W} becomes exact.
The resulting static transport coefficient $\sigma_Q(\omega=0,\mu)$ is plotted as a function of $\mu/T$ in Fig.~\ref{f:sigmaQ_of_mu}.
It is interesting to note that in the regime $\mu\gg T$, $\sigma_Q$ decays as $(T/\mu)^{2}$, reflecting that relativistic physics and quantum criticality associated with the presence of particles and antiparticles becomes less and less relevant.\begin{figure}
\includegraphics[width=0.5\textwidth]{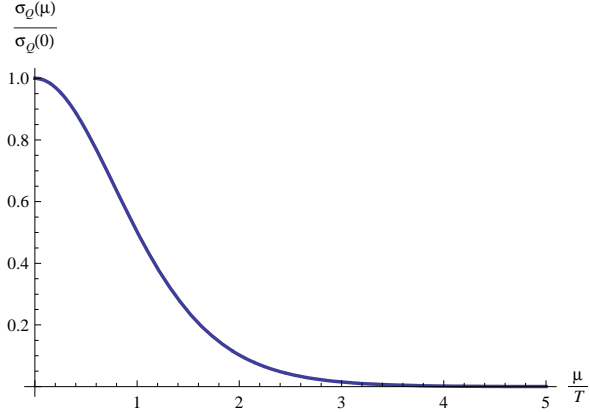}
\caption{The normalized transport coefficient $\sigma_Q(\mu,\omega=0)$ as a function of $\mu/T$.}
\label{f:sigmaQ_of_mu}
\end{figure}

The above discussion refers to the typical relaxation time for modes $g(k)$ which vary significantly over a range of order $T$ around $k=\mu$. These modes are associated with so-called ``vertical processes" that degrade thermal currents very efficiently~\cite{ziman}. However, their contribution to the electrical current is small, which is reflected by the smallness of $\sigma_Q$ scaling as $(T/\mu)^2$. The electrical conductivity is thus largely dominated by the first term in (\ref{sigmapure}), and this holds true also in the presence of weak disorder, see Eq.~\eqref{eq:siwd} below. This statement is independent of whether electron-electron interactions are screened. However, the same is not true for the thermal conductivity which turns out to be sensitive to unscreened Coulomb interactions even when $\mu\gg T$. This will be discussed further in Sec.~\ref{s:WF}.

\subsection{Quantum critical regime}
Note that close to particle-hole symmetry, $\mu\ll T$, $\tau_{\rm ee}^{-1}$ is essentially given by the temperature, $\tau_{\rm ee}^{-1}\sim \alpha^2 k_BT/\hbar$, cf.~Eq.~(\ref{tauee}). This is a hallmark of many quantum critical systems where often a relativistic effective theory emerges (as characterized by a dynamical critical exponent $z=1$)~\cite{ssbook} with the temperature $T$ being the only energy scale left.

At particle-hole symmetry, the above expressions (\ref{sigmapure},\ref{sigmaQmu}) become identical to the results reported  in a previous publication~\cite{FSMS} for the clean limit:
\begin{eqnarray}
\label{sigma0}
\sigma_{xx}(\omega,\mu=0)=\sigma_Q^0 \frac{1}{1-i\omega \tau^{0}_{\rm ee}}\,.
\eea
In the limit $\eta\to 0$ one finds the explicit numerical values
\bea
\sigma_Q^0&=&\frac{1}{\alpha^2}\frac{e^2}{\hbar}\frac{N}{2}\hat g_1(0)\left(\frac{\log(2)}{\pi}\right)^2\nn\\
&=& \frac{0.121}{\alpha^2} \frac{e^2}{\hbar}=\frac{0.760}{\alpha^2} \frac{e^2}{h}\,,\\
\tau^0_{\rm ee}&=&\frac{\hbar}{\alpha^2k_BT}\frac{\hat g_1(0)\log(2)}{2\pi}= 0.274\frac{\hbar}{\alpha^2k_BT}\,,
\eea
where we have used $\hat g_1(0)\approx 1.24$.
Accordingly, the inelastic mean free path evaluates to
\bea
\label{mfpath}
\ell=v_F\tau^0_{\rm ee}&=& \frac{2.3}{\alpha^2 T\rm{[K]}}\,\mu{\rm m}\,.
\eea

\subsection{Weak disorder}
In a clean system, the denominator in the expression (\ref{cleanMinverse}) for the conductivity vanishes as $\omega\to 0$ since $M_{0}, M_1\sim \omega$. This reflects momentum conservation due to which the mode $\phi_0$ does not decay.

However, the presence of disorder shifts the pole from $\omega=0$ into the negative halfplane, such that the denominator behaves as $\omega+i\tau_{\rm imp}^{-1}$. This defines the elastic scattering rate $\tau_{\rm imp}^{-1}$.
To leading order in weak disorder it is given by
\bea
\label{tauimp}
\tau_{\rm imp}^{-1}&=&\frac{\Delta (\rho^++\rho^-)}{\eP}=\frac{\rho_{\rm imp}}{\hbar}\left(\frac{\pi Ze^2}{\epsilon_r}\right)^2 \frac{\rho^++\rho^-}{\eP}\nn\\
&\sim& \frac{\rho_{\rm imp}}{\hbar}\left(\frac{\pi Ze^2}{\epsilon_r}\right)^2 \frac{1}{{\rm max}[k_B T,\mu]}\,.
\eea

One can explicitly verify that in an expansion in weak disorder, i.e., in $\tau^{-1}_{\rm imp}/\tau^{-1}_{\rm ee}\sim \Delta/\alpha^2$, the conductivity is given by
\begin{eqnarray}
\label{eq:siwd}
&&\sigma_{xx}(\omega;\mu,\Delta)=
 \frac{e^2}{\tau_{\rm imp}^{-1}-i\omega}\frac{\rho^2v_F^2}{\eP}+\sigma_Q+\delta \sigma(\Delta, \omega,\mu)\;,\nonumber \\
\end{eqnarray}
with $\delta \sigma(\Delta, \omega,\mu)={\cal O}(\Delta/\alpha^2)$.

Similarly one finds for the other thermo-electric response functions
\begin{eqnarray}
\label{eq:fullalpha_kappa}
\alpha_{xx}(\omega;\mu,\Delta)&=&
\frac{e}{\tau_{\rm imp}^{-1}-i\omega}\frac{s\rho v_F^2}{\eP}-\frac{\sigma_Q}{e}\frac{\mu}{T}+\delta \alpha(\Delta, \omega,\mu),\nn\\
\overline{\kappa}_{xx}(\omega;\mu,\Delta)&=&
\frac{1}{\tau_{\rm imp}^{-1}-i\omega}\frac{s^2 T v_F^2}{\eP}+\frac{\sigma_Q}{e^2}\frac{\mu^2}{T}+\delta \kappa(\Delta, \omega,\mu),\nn
\end{eqnarray}
with disorder corrections of order $\delta \alpha,\delta \overline{\kappa} = {\cal O}(\Delta/\alpha^2)$. Remarkably, by dropping these higher order terms, we recover precisely the expressions predicted by relativistic hydrodynamics with a phenomenological momentum relaxation rate implemented via $\omega\to \omega+i\tau^{-1}_{\rm imp}$.
Since disorder breaks the relativistic invariance of the particle-hole plasma, it is not surprising that it eventually spoils the relativistic structure of the response functions when the disorder-induced elastic scattering rate $\tau_{\rm imp}^{-1}$ becomes comparable to $\tau_{\rm ee}^{-1}$.

\section{Crossover from quantum critical to 'Fermi-liquid' regime at $\mu\gg T$}
\label{s:Fermiliquid}
\subsection{Two mode approximation}\label{sec:tma}
As mentioned before, in the Boltzmann approach we are not restricted to small disorder satisfying $\Delta/\alpha^2\ll 1$. In order to obtain insight into the crossover from the hydrodynamic to the disorder dominated regime, we will adopt here the two mode approximation. The latter is not only convenient for analytical treatments, but also turns out to describe both the interaction and disorder dominated limit well.

In a regime dominated by elastic scattering one usually resorts to the relaxation time approximation.
\bea
g_\lambda(k)= -(\partial f^0/\partial E) \tau_k \mathbf{v}_\mathbf{k} e\mathbf{E}.
\eea
In the case of unscreened Coulomb scatterers this is exact if the relaxation time is chosen to increase linearly with the energy, $\tau_k\sim k$.
In the 'non-relativistic' limit $\mu\gg T$
 where thermal excitations of only electrons {\em or} holes are relevant, this is essentially equivalent to making an Ansatz with the single mode $\phi_1$, and the corresponding relaxation time is related to the coefficient $\psi_{1}$ in (\ref{basisexpansion}) by
\bea
\tau_k=|\psi_1| \frac{\hbar v_F k}{k_BT} \frac{\hbar}{k_B T}\,.
\eea

On the other hand, in the regime where electron-electron scattering dominates we have already argued that enhanced forward scattering allows us to restrict to the two modes $\phi_{0,1}$ to leading order in $1/\log(\eta)$. This suggests that an approximate description of the Boltzmann equation~\eqref{trans0} restricted to those two modes actually yields a rather accurate description of the whole crossover from quantum critical to disorder dominated regime.

In this approximation the full expression for the longitudinal conductivity assumes the form:
\begin{eqnarray}
\label{eq:fullsigma}
&&\sigma_{xx}(\omega;\mu,\Delta)=
e^2 \frac{1}{\tau_{\rm imp}^{-1}-i{\omega}}\frac{\rho^2v_F^2}{\eP}+\sigma_{xx}'\\
&&\sigma_{xx}'=e^2\frac{\left [I_+^{(1)}-\frac{\rho[\Delta I_-^{(1)}-i{\omega}\rho]^2}{\Delta I_+^{(2)} -i{\omega} (\eP)}\right]^2}{\frac{N}{2}(\frac{1}{g_1}+\frac{\Delta}{2\pi})-i{\omega}I_+^{(1)}  -\frac{[\Delta I_-^{(1)}-i{\omega}\rho]}{\Delta I_+^{(2)} -i{\omega} (\eP)}},\nn
\end{eqnarray}
with $\tau^{-1}_{\rm imp}$ as given in (\ref{tauimp}) and the functions $I_{\pm}^{(k)}(\mu)$ defined in App.~\ref{app:Melements}. In the second term we have set the factors of $\hbar=v_F=k_BT=1$.
We have decomposed the response function into two parts:
The first term is independent of the inelastic scattering rate (that is, of $\hat g_1$), and is thus entirely determined by impurity scattering.
The second term, $\sigma_{xx}'$,
has a finite d.c. limit as $\Delta\to 0$, and it reduces to the quantum critical value of conductance of pure undoped graphene as $\rho,\mu\to 0$. However, the first term diverges in the clean static limit except at the particle-hole symmetric point where its numerator vanishes.
\begin{figure}
\includegraphics[width=0.45\textwidth]{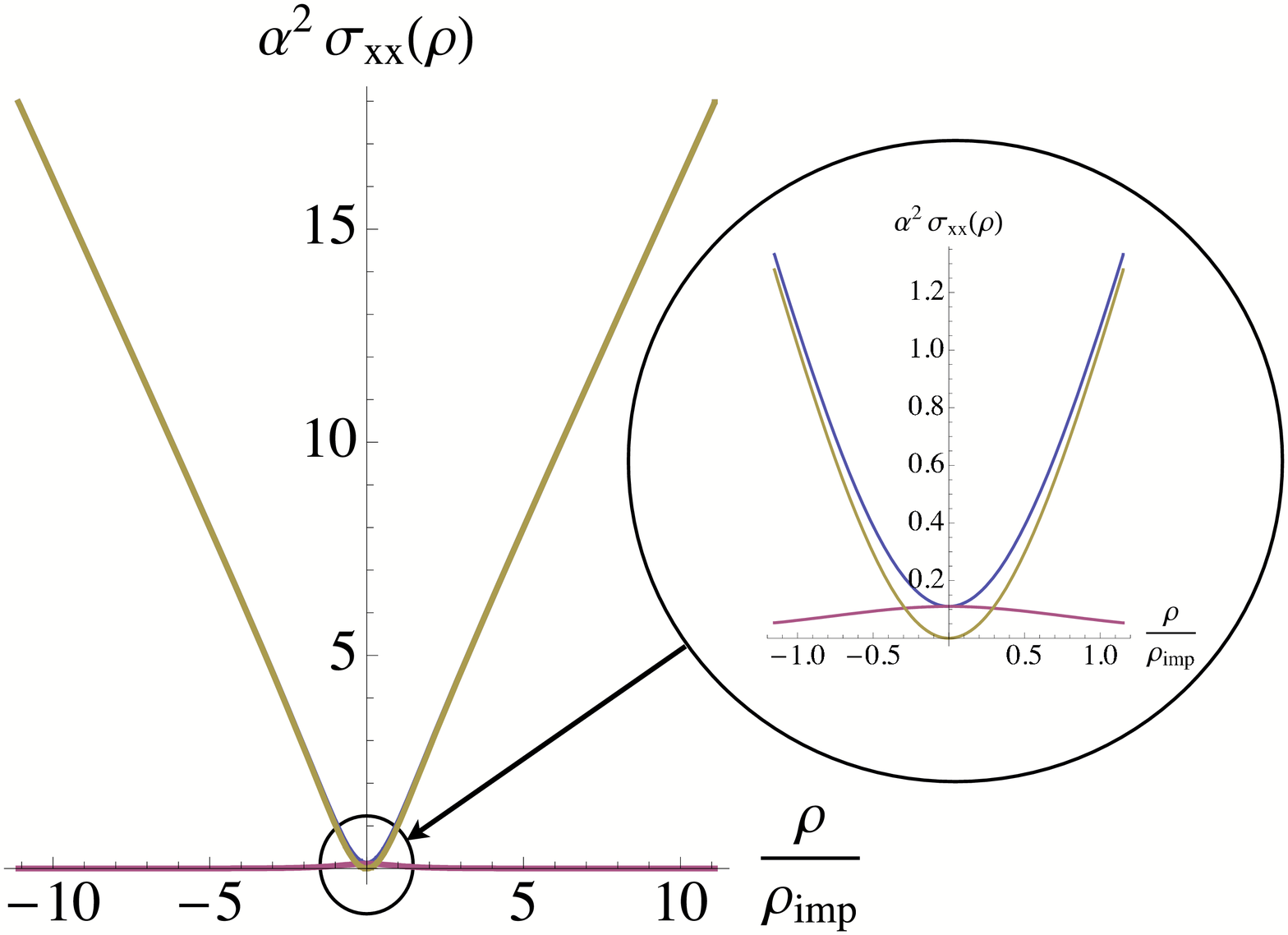}
\caption{The d.c. conductivity, $\alpha^2\sigma_{xx}$ in units of $e^2/\hbar$, as a function of the ratio of doped carrier density to impurity density. The disorder strength and temperature have been chosen such that $\Delta/\alpha^2=0.25$. The plot shows the crossover from the quantum critical regime, where the conductivity is essentially limited by inelastic scattering between electrons, to the regime dominated by elastic scattering from Coulomb impurities at higher doping. In the latter regime the conductivity increases linearly with doped carrier density, see Eq.~(\ref{sigmaproptorho}). The red and yellow curves correspond to the two terms in Eq.~(\ref{eq:fullsigma}), respectively: the contribution from modes relaxing due to inelastic scattering, and the contribution from the momentum mode, which is only disorder limited and dominates at large density.
}
\label{f:sigma_of_n}
\end{figure}

At zero doping the above two-mode approximation reduces to the same expression as (\ref{sigma0}), with the replacement $\alpha^2/\hat g_1\to \alpha^2/\hat g_1 +\Delta/(2\pi)$. This means that an impurity scattering rate proportional to $\Delta$ is added to the inelastic scattering rate in the denominator.
The purely interaction dominated, 'quantum-critical' conductivity is thus visible only in weak disorder where $(\Delta/\alpha^2) \hat g_1/(2 \pi)\ll 1$; otherwise disorder dominates the response at all dopings. The crossover from interaction dominated to disorder dominated transport occurs when
the two terms in (\ref{eq:fullsigma}) are approximately equal, i.e., when $\rho\sim \rho_{\rm imp}$, cf.~Fig.~\ref{f:sigma_of_n}.

\subsection{Linear conductivity at large doping}

We can make direct contact with earlier studies which considered disorder dominated transport in the non-interacting Fermi liquid regime of graphene. From Eq.~\eqref{eq:fullsigma} and Eq.~\eqref{eq:Fermi} it is easy to check that at large $\mu\gg T$, the first term in Eq.~\eqref{eq:fullsigma} dominates the static conductivity which is thus entirely determined by impurity scattering.
The full conductivity and the contributions of either term are plotted in Fig.~\ref{f:sigma_of_n}.

In the disorder dominated limit we recover the expression for the conductivity of non-interacting electrons in doped graphene in the presence of Coulomb impurities, previously reported in Ref.~\onlinecite{rmp}
\begin{eqnarray}
\sigma_{xx}(\omega=0;\mu\gg T) &\approx& \frac{e^2\rho^2v_F^2\tau_{\rm imp}}{\eP}\nn\\
&=& \frac{2}{\pi}\frac{1}{(Z \alpha)^2}\frac{e^2}{h}\frac{\rho}{\rho_{\rm imp}}\,.
\label{sigmaproptorho}
\eea
Note that in our approximation where we neglect screening of the impurities, there is no constant offset to the term linear in $\rho$.
It would be straightforward to include RPA screening in the impurity potential, which  is known to modify the numerical prefactor of the linear density dependence~\cite{Nomura,DasSarmaGalitski,AdamSarma} in (\ref{sigmaproptorho}) and produce a positive offset proportional $\alpha^2$, see Ref.~\cite{Trushin}

\subsection{Mott's law}
\label{Mott}
In the regime $\mu\gg T$ we expect to recover ordinary Fermi liquid behavior.
In the approximation with two modes $\phi_{0,1}$, the thermopower $\alpha_{xx}$ is given by
\bea
\label{alphatwomode}
&&\alpha_{xx}(\mu,\omega=0)= \frac{e s\rho v_F^2 \tau_{\rm imp}}{\eP}+\alpha_{xx}'\\
&&\alpha_{xx}'= -\left(I^{(1)}_+ -\frac{I^{(1)}_- \rho}{I^{(2)}_+}\right)
\frac{s I^{(1)}_-  + I^{(2)}_+(\mu I^{(1)}_+ -\rho)}{I^{(2)}_+/g_1 - \Delta\left[I_-^{(1)}\right]^2},\nn
\eea
where in the second term we have again dropped factors of $\hbar,v_F,k_B$ and $T$.
At large $\mu\gg T$, the first term dominates again, similarly as for the conductivity.

In a Fermi liquid where scattering at low temperature is dominated by impurities, one expects the thermopower to be related to the conductivity according to Mott's law,
\bea
\alpha_{xx}(\mu,\omega=0) =-\frac{\pi^2}{3e}k_B^2 T\frac{d\sigma(\mu,\omega=0)}{d\mu}.
\eea
Indeed, using Sommerfeld expansions for $\rho$ and $s$ in the above results, one can easily show that this relation holds for $\mu\gg T$. Not unexpectedly, however it fails in the quantum critical regime $\mu\lesssim T$.
The ratio $-\frac{3e k_B^2 T}{\pi^2}\frac{\alpha_{xx}(\mu)}{d\sigma/d\mu(\mu)}$ is plotted in Fig.~\ref{fig:Mott} as a function of $\mu/T$ (assuming unscreened electron-electron interactions).
Note that $\alpha_{xx}$ tends to zero as $\mu$ vanishes. The same holds for $d\sigma/d\mu(\mu)$ but the ratio of the two quantities remains finite.
\begin{figure}
\includegraphics[width=0.45\textwidth]{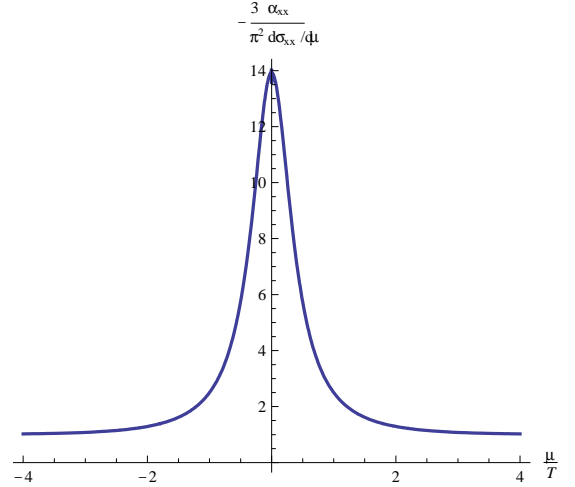}
\caption{The ratio $R\equiv -\frac{3e k_B^2 T}{\pi^2}\frac{\alpha_{xx}(\mu)}{d\sigma/d\mu(\mu)}$ as a function of $\mu/T$. In the Fermi liquid regime ($\mu\gg T$) $R$ tends to $1$ as predicted by Mott's law.
}
\label{fig:Mott}
\end{figure}

\subsection{Wiedemann-Franz law}
\label{s:WF}
In the clean limit, the Lorentz ratio
\bea
\label{Lorentz}
L=\frac{\kappa_{xx}}{\sigma_{xx}},
\eea
diverges as $\mu\to 0$ instead of acquiring the standard Fermi liquid value of $\frac{k_B^2T}{e^2}\frac{\pi^2}{3}$. This is another peculiarity of the quantum critical regime.

One would expect to recover the Fermi liquid value of the Lorentz ratio, or in other words, the Wiedemann Franz law, by going to sufficiently low temperatures in the Fermi liquid regime $\mu\gg T$. This is indeed so if we assume the electron-electron interactions to be screened, such that inelastic scattering processes degrading the energy current become inefficient as compared to impurity scattering at low $T$. This is also in qualitative agreement with the thorough analysis of Ref.~\onlinecite{Catelani}, which in addition discusses the contribution of bosonic particle-hole excitations to thermal transport.

If the interactions are not screened, however, the Lorentz ratio remains sensitive to them and does not reduce to the Fermi liquid value. The reason is as follows. Even though an expression analogous to (\ref{eq:fullsigma},\ref{alphatwomode}) exists for $\overline{\kappa}_{xx}$, and is dominated by its first term $s^2Tv_F^2\tau_{\rm imp}/(\eP)$, the same is not true for the thermal conductivity $\kappa_{xx}=\overline{\kappa}_{xx}-T\alpha_{xx}^2/\sigma_{xx}$ which is sensitive to the subleading corrections to the dominant terms in $\sigma_{xx},\alpha_{xx}$ and $\overline{\kappa}_{xx}$.
If the electron-electron interactions are not screened, implying that $\hbar\tau^{-1}_{\rm ee}(\mu)\sim \mu$ for large $\mu$, the effects of these interactions persist in these subleading terms even deep in the Fermi liquid regime $\mu\gg T$.


\section{Magnetotransport}
\label{s:finiteB}
In this section we turn to transport in the presence of a magnetic field. The matrix formalism with the appropriate choice of basis functions will prove particularly useful to recover the hydrodynamic response and to find explicit expressions for the more general structure of the thermo-electric response functions.
To this end, we need to develop the matrix formalism of Sec.~\ref{s:Boltzmann}. We anticipate that the dynamics of the momentum mode $\phi_0$ captures most of the physics in the hydrodynamic regime. It is therefore convenient to introduce a notation that singles out the components pertaining to it:
\begin{eqnarray}
\vec{\psi}_{||,\perp}=\left (\begin{array}  {c} \psi^0_{||,\perp} \\ \vec{\psi}'_{||,\perp} \end{array} \right)\, ; \quad \mathcal{F}^D=\left ( \begin{array}{c} F_0^D \\ \vec{F}^D  \end{array} \right)\,,
\end{eqnarray}
where $D=E,T$.
We recall that our basis choice implies $\vec{F}^D=(F_1^D,0,\dots,0)^T$. Further, for the matrices in the Boltzmann equation we have
\begin{eqnarray}
\mathcal{M}=\left (\begin{array}{cc} M_0 & \vec{M}_1^T\\ \vec{M}_1 & {\bf{M}} \end{array} \right)\, ; \quad \mathcal{B}=\left (\begin{array}{cc} B_0 & \vec{B}_1^T\\ \vec{B}_1 & {\bf{B}} \end{array} \right)\,.
\end{eqnarray}
It proves convenient to define a $2\times 2$ matrix
\begin{eqnarray}
\label{R0}
{R}_0 &=& \left ( \begin{array}{cc}  M_0 & -B_0 \\ B_0  & M_0  \end{array}\right)
\eea
as well as $2\times (2{\cal N})$ and $(2{\cal N})\times 2$ matrices ${\cal R}_1,\overline{{\cal R}}_1$,
\bea
\mathcal{R}_1 = \left ( \begin{array}{cc} \vec{M}^T_1 & -\vec{B}^T_1\\ \vec{B}^T_1& \vec{M}^T_1  \end{array}\right)\,;\quad
\overline{{\cal R}}_1 = \left ( \begin{array}{cc} \vec{M}_1 & -\vec{B}_1\\ \vec{B}_1& \vec{M}_1  \end{array}\right)\,.
\end{eqnarray}
This allows us to reformulate the matrix equation, splitting it into a zero component,
\begin{eqnarray}
\label{zeroEq}
{R}_0 \left (\begin{array}{c} \psi^0_{||}\\ \psi^0_\perp \end{array} \right) +\mathcal{R}_1 \left (\begin{array}{c} \vec{\psi}'_{||}\\ \vec{\psi}'_\perp \end{array} \right)= \left (\begin{array}{c} F_0^D\\ 0 \end{array} \right)\,,
\end{eqnarray}
and a vector component,
\begin{eqnarray}
\left(\begin{array}{cc} {\bf{M}}& -{\bf{B}} \\ {\bf{B}}& {\bf{M}} \end{array} \right)\left(\begin{array} {c}\vec{\psi}'_{||} \\ \vec{\psi}'_{\perp}\end{array}\right)=\left ( \begin{array}{c}  \vec{F}^D \\ \vec{0}\end{array}\right)-\overline{{\cal R}}_1 \left(\begin{array} {c}\psi^0_{||} \\ \psi^0_\perp \end{array}\right)\,.
\end{eqnarray}
Defining
\begin{eqnarray}
\mathcal{G}=\left(\begin{array}{cc} {\bf{M}}& -{\bf{B}} \\ {\bf{B}}& {\bf{M}} \end{array} \right)^{-1}
\end{eqnarray}
we can formally solve for the vector
\begin{eqnarray}
\left (\begin{array} {c}\vec{\psi}'_{||} \\ \vec{\psi}'_\perp \end{array} \right) = \mathcal{G} \left [\left (\begin{array}{c} \vec{F}^D \\ 0 \end{array} \right) - \overline{{\cal R}}_1 \left( \begin{array}{c} \psi^0_{||}\\ \psi^0_\perp \end{array}\right) \right]\,.
\end{eqnarray}
Inserting in (\ref{zeroEq}) and
solving for the zero mode components leaves us with
\begin{eqnarray}
\left(\begin{array}{c} \psi_{||}^0  \\ \psi_\perp^0 \end{array} \right) = \left [ {R}_0 - \mathcal{R}_1 \mathcal{G} \overline{{\cal R}}_1 \right]^{-1} \left [ \left (\begin{array} {c} F^D_0 \\ 0 \end{array}  \right) - \mathcal{R}_1 \mathcal{G}\left (\begin{array} {c} \vec{F}^D\\ \vec{0}\end{array} \right)\right]\,.
\end{eqnarray}
The response functions can finally be calculated from the solution for $\psi_{\parallel,\perp}^{0,1}$.

\subsection{Magnetotransport in the absence of disorder}
Significant progress can be made due to a great simplification which occurs in the above formulae if there is no disorder.
As we have shown in (\ref{M0n}), in this case the vectors $\vec{M}_1$ and $\vec{B}_1$ only have one non-vanishing component, $\vec{M}_1=M_1\vec{e}_1$ and $\vec{B}_1=B_1\vec{e}_1$. The same holds in general for the driving terms, $\vec{F}^D=F^D_1\vec{e}_1$. One can then easily convince oneself that
one only needs to know the $2\times2$ matrix
\bea
G\equiv \left( \begin{array}{c}
\vec{e}^T_1  \\
\vec{e}^T_1  \end{array}\right) {\cal G}
\left( \begin{array}{cc}
\vec{e}_1  & \vec{e}_1
\end{array}\right) \equiv
\left( \begin{array}{cc}
g_1  & g_2\\
-g_2 & g_1
\end{array}\right)
\eea
where
\bea
g_1(\omega,B)&\equiv&\mathbf{e}_1^T\cdot\left({\mathbf M} +{\mathbf B}{\mathbf M}^{-1}{\mathbf B}\right)^{-1}\mathbf{e}_1,\\
g_2(\omega,B)&\equiv&\mathbf{e}_1^T\cdot{\mathbf M}^{-1}{\mathbf B}\left({\mathbf M} +{\mathbf B}{\mathbf M}^{-1}{\mathbf B}\right)^{-1}\mathbf{e}_1,\nn
\eea
and  the $2\times 2$ matrix
\bea
\label{R1}
R_1 &=& \left ( \begin{array}{cc} M_1 & -B_1\\ B_1& M_1  \end{array}\right) \\
&=& \frac{2}{N} \left ( \begin{array}{cc} -i\omega\rho & -BI_+^{(1)}(\mu)\\ BI_+^{(1)}(\mu)& -i\omega\rho  \end{array}\right)\,,\nn
\eea
both acting in $(\parallel,\perp)$-space. $I_+^{(1)}$ has been defined in App.~\ref{app:Melements}. Further note that for $B=0$, $g_1$ coincides with the transport coefficient defined in the previous sections.
For convenience we give the explicit form of $R_0$ (\ref{R0}) using results from App.~\ref{app:Melements}:
\bea
\label{R0_2}
R_0 =  \frac{2}{N} \left ( \begin{array}{cc} -i\omega(\eP) & -B \rho\\ B \rho& -i\omega(\eP)  \end{array}\right)\,.
\eea
The above equations immediately yield the solution for the $0$ and $1$ components of $\vec{\psi}_{\parallel,\perp}$,
\begin{eqnarray}
\label{01Equation_1}
\left(\begin{array}{c} \psi_{||}^0  \\ \psi_\perp^0 \end{array} \right) &=& \left [ {R}_0 - R_1 G  R_1 \right]^{-1} \left [ \left (\begin{array} {c} F^D_0 \\ 0 \end{array}  \right) - R_1 G\left (\begin{array} {c} F_1^D\\ 0\end{array} \right)\right]\,, \nonumber \\
\left (\begin{array} {c} \psi^1_{||} \\ \psi^1_\perp \end{array} \right) &=& G \left [\left (\begin{array}{c} F_1^D \\ 0 \end{array} \right) - R_1 \left( \begin{array}{c} \psi^0_{||}\\ \psi^0_\perp \end{array}\right) \right]\,.
\end{eqnarray}

It is a simple matter to express all response coefficients using (\ref{currents}). For the thermal conductivity, e.g., one finds
\begin{widetext}
\bea
\label{kappafull}
&&\overline{\kappa}_{xx}=
\frac{N}{2} \left( \begin{array}{cc}
F_0^T & F_1^T\end{array}\right) \cdot \left( \begin{array}{cc}
\left[({R}_0-{R}_1{G}{R}_1)^{-1}\right]_{\parallel,\parallel} &
-\left[({ R}_0-{R}_1{G}{R}_1)^{-1}{R}_1{G}\right]_{\parallel,\parallel} \\
-\left[{G}{R}_1({R}_0-{R}_1{G}{R}_1)^{-1}\right]_{\parallel,\parallel} &
\left[{G}+{G}{R}_1({R}_0-{R}_1{G}{R}_1)^{-1}{R}_1 {\cal G}\right]_{\parallel,\parallel}
\end{array}\right)
\left( \begin{array}{c}
F_0^T \\ F_1^T\end{array}\right),\nn
\eea
\bea
&&\overline{\kappa}_{xy}=
\frac{N}{2} \left( \begin{array}{cc}
F_0^T & F_1^T\end{array}\right) \cdot \left( \begin{array}{cc}
\left[({R}_0-{R}_1{G}{R}_1)^{-1}\right]_{\parallel,\perp} &
-\left[({ R}_0-{R}_1{G}{R}_1)^{-1}{R}_1{G}\right]_{\parallel,\perp} \\
-\left[{G}{R}_1({R}_0-{R}_1{G}{R}_1)^{-1}\right]_{\parallel,\perp} &
\left[{G}+{G}{R}_1({R}_0-{R}_1{G}{R}_1)^{-1}{R}_1 {\cal G}\right]_{\parallel,\perp}
\end{array}\right)
\left( \begin{array}{c}
F_0^T \\ F_1^T\end{array}\right).
\eea
\end{widetext}

These expressions are further analyzed below.

\subsection{D.C. response of pure samples in a magnetic field}

\subsubsection{Small fields - Hydrodynamic regime}
The hydrodynamic regime corresponds to the limit $\tau_{\rm ee}^{-1}\gg \omega,\omega_c$, where one may approximate
\bea
g_1&\approx& [\mathbf{M}^{-1}]_{11}+{\cal O}([\omega_c^{\rm typ}\tau_{\rm ee}]^2),\nn\\
g_2&\approx& {\cal O}(\omega^{\rm typ}_c\tau_{\rm ee}).
\eea
We will see that corrections to higher order in the magnetic field correspond to corrections beyond the hydrodynamic analysis, as the latter indeed relies on the smallness of the magnetic field. In Eq.~(\ref{smallB}) below we will give a quantitative criterion for the onset of corrections in large fields.

At finite magnetic field and in the absence of disorder, the Boltzmann transport theory predicts a vanishing d.c. conductivity and thermopower,
\bea
\sigma_{xx}(\omega=0)=\alpha_{xx}(\omega=0)=0.
\eea
This can be understood as a consequence of Lorentz invariance: In a reference frame moving at the constant velocity $\mathbf{ v}_D=\mathbf{E}\times \mathbf{B}/B^2$ with respect to the lab frame, the observed electric field vanishes, and hence, in that frame all currents vanish. Upon transforming back, since $\mathbf{ v}_D$ is perpendicular to $\mathbf{E}$, this implies also a vanishing longitudinal response to the electric field in the lab frame. The transverse d.c. response takes an equally simple form: It yields the standard Hall effect, $\sigma_{xy}=\rho/B$, and the transverse Peltier effect $\alpha_{xy}=s/B$, which can be interpreted as charge and entropy density drifting with the velocity $\mathbf{v}_D$ ~\cite{bhaseen}. These results are in agreement with the hydrodynamic description of Sec.~\ref{MHD}, but hold much more generally due to Lorentz invariance, even when $\omega^{\rm typ}_c \tau_{\rm ee}\gg 1$.

In contrast to the electrical conductivity and the thermopower, the longitudinal thermal conductivity of clean, interacting Dirac particles remains finite in the d.c. limit. From the formula (\ref{kappafull}) and the matrices (\ref{R1},\ref{R0_2}) one finds after some algebra the result
\bea
\overline{\kappa}_{xx}=\frac{(\eP)^2\sigma_Q}{\rho^2+B^2[I_+^{(1)}]^4(g_1^2+g_2^2)-2\rho g_2 B[I_+^{(1)}]^2}\,.\nn
\eea
The expression simplifies further at particle hole symmetry where $\rho=0$ and the matrix $G$ is diagonal $(g_2=0)$. This is a consequence of ${\bf M}$ and ${\bf B}$ being symmetric and antisymmetric with respect to $\lambda\to -\lambda$, respectively. $g_2$ vanishes since it is an expectation value of a matrix under antisymmetric under $\lambda \to -\lambda$, evaluated on the mode $\phi_1$ which has definite symmetry under the same transformation.

Using $\sigma_Q(\mu=0)=g_1[I_+^{(1)}]^2|_{\mu=0}= 0.156 e^2/\hbar\alpha^2$ (for $\eta=0.01$), we find
\bea
\label{kappaMHDrecover}
&&\kappa_{xx}(\omega=0, \mu=0)=\overline{\kappa}_{xx}(\omega=0, \mu=0)\\ &&\quad\quad=\frac{1}{B^2}\frac{s^2T}{\sigma_Q(0)} = C_{\rm MHD}\alpha^2  \left(\frac{T^2}{eB v_F^2}\right)^2\, k_B^2T,\nn
\eea
with
\bea
\label{CMHD}
C_{\rm MHD}=\frac{1}{0.156}\left(\frac{9\zeta(3)}{\pi}\right)^2=76.0.
\eea
Note that the expression (\ref{kappaMHDrecover}) is precisely the prediction of magnetohydrodynamics, cf. Eq.~(\ref{kappaMHD}).

The transverse response at particle-hole symmetry is found to vanish in a closed circuit,
\bea
\overline{\kappa}_{xy}(\omega=0, \mu \to 0)&=&0,
\eea
in agreement with hydrodynamics as well.
However, $\kappa_{xy}$ diverges, since
\bea
\kappa_{xy}(\omega=0, \mu=0)&=& -\frac{\alpha_{xy}^2}{\sigma_{xy}}\sim \frac{B}{\rho}\to \infty.
\eea

Note that, similarly as in the hydrodynamic approach, the limits $B\to 0$ and $\omega \to 0$ do not commute. That is, the small field limit of the d.c. response discussed above does not correspond to the d.c. limit of the $B=0$ response. Indeed, the latter diverges in general as $1/\omega$ due to momentum conservation except at zero doping, whereas any finite field $B$ leads to a vanishing d.c. conductivity in clean systems.

\subsubsection{Large fields - Ballistic regime}

The Boltzmann approach allows us to go beyond the hydrodynamic regime. In particular, it is interesting to study the large field limit, and how the crossover, as controlled by the parameter $b/\alpha^2$, takes place. The static response in the regime where this parameter is small and the inelastic scattering time is the shortest timescale, was discussed above. Further, in the subsequent  section we will show  that
the complete low frequency response is well described by relativistic magnetohydrodynamics, too.

The regime of large fields can be addressed by determining the deviations from equilibrium in a perturbative expansion in $\alpha^2/b$.
Such an approach was taken in Ref.~\onlinecite{bhaseen} where the thermoelectric response of relativistic bosons was studied. Those are the low energy quasiparticles expected at the transition from a commensurate Mott insulator to a superfluid, for which the simplest model is the Wilson Fisher fixed point occurring in a $\phi^4$ theory.
In an $\epsilon=3-d$ expansion~\cite{damless,ssbook} around $d=3$, the fixed point value of the quartic coupling scales as $\epsilon$, which takes the role of $\alpha$ in our case. In the spirit of an $\epsilon$ expansion, the authors of Ref.~\onlinecite{bhaseen} assumed $\epsilon$ to be small and thus focussed on the  'ballistic' limit of magnetoresponse where the parameter $b/\epsilon^2$ is large. Here we analyze the large field limit in an analogous way, and work out the crossover to the hydrodynamic low field regime.

We recall that the field should still be weak enough to avoid the effects of Landau quantization. This requires that the typical cyclotron energy, $\hbar\omega_c^{\rm typ}$ be smaller than $k_BT$, or equivalently, $b\ll1$.

In the regime, $\alpha^2\ll b\ll 1$ the transverse deviation of the distribution functions leads to a transverse current whose deflection by the magnetic field balances the current-driving tendency of the applied fields. The longitudinal currents are smaller by an extra factor of $\alpha^2/b$. Formally, to leading order, the solution of the Boltzmann equation~\eqref{trans0} is given by
\bea
\label{largeBperp}
\vec{\psi}_\perp&\approx &-{\cal B}^{-1}{\cal F},\\
\label{largeBpara}
\vec{\psi}_\parallel &\approx & -{\cal B}^{-1}{\cal M}^{\rm Cb}\vec{\psi}_\perp={\cal B}^{-1}{\cal M}^{\rm Cb}{\cal B}^{-1}{\cal F}.
\eea

Let us focus on the thermal response again. Writing the driving thermal gradient as
\bea
{\cal F}(k)=-f_{\lambda k}^0[1-f^0_{\lambda k}] g_T(\lambda, k) \nabla T \cdot \mathbf{e_k}
\eea
where $g_T(\lambda,k)=k-\lambda \mu$ we find from (\ref{largeBperp})
\bea
\label{gperp}
g_\perp(\lambda, k)=-\frac{\lambda k}{b}g_T(\lambda, k)\,|\nabla T|.
\eea
The longitudinal component follows form the balance between the deflection of the longitudinal current and the inelastic collisions which degrade the transverse currents
\bea
g_\parallel(\lambda,k)f_{\lambda k}^0[1-f^0_{\lambda k}]=-\frac{\lambda k}{b}\left[{\cal M}^{\rm Cb}g_\perp\right](\lambda,k).
\eea
We focus on the particle-hole symmetric case where from (\ref{gperp}) $g_\perp(\lambda, k)=\lambda k^2\,(-\nabla T)/b$. The longitudinal thermal conductivity $\kappa_{xx}$ turns out to scale parametrically in the same way as in the hydrodynamic regime, however with a different numerical prefactor:
\bea
\overline{\kappa}_{xx}(\omega=0,\mu=0)&=&\kappa_{xx}(\omega=0,\mu=0)\\
&=&C_{b\gg \alpha^2} \alpha^2  \left(\frac{T^2}{eB v_F^2}\right)^2\, k_B^2T.\nn
\eea
Here,
\bea
\label{ClargeB}
&&C_{b\gg \alpha^2}=\frac{N}{2} \tilde{g} \cdot{\cal M}^{\rm Cb}\tilde{g}\\
&&\quad\equiv \frac{N}{2} \sum_\lambda\int \frac{d^2k}{(2\pi)^2} \tilde{g}(\lambda,k){\cal I}_{\rm coll}^{\rm Cb}[\lambda,k\,|\{\tilde{g}\}] = 383.98,\nn
\eea
where $\tilde{g}(\lambda,k)=\lambda k^2$, and we have regularized the collision term by fixing the RPA screening to $\eta=0.01$.
The crossover from the magnetohydrodynamic regime to the regime of large fields has been calculated using the full matrix formalism of the previous section with 12 basis functions, and the result is plotted in
Fig.~\ref{fig:kappacrossover}. Notice that the crossover extends over rather large fields before the asymptotic limit~(\ref{ClargeB}) is reached.

\begin{figure}
\includegraphics[width=0.45\textwidth]{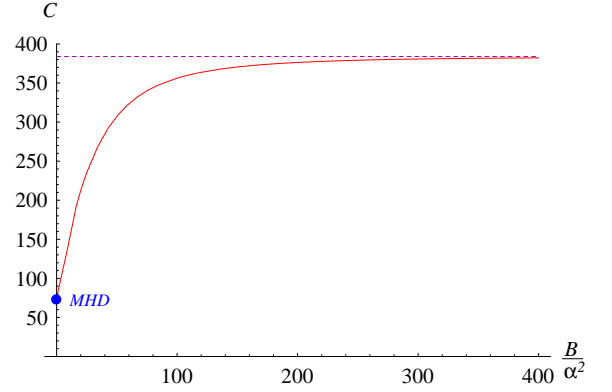}
\caption{The thermal conductivity at particle-hole symmetry scales as $\kappa_{xx}\sim B^{-2}$ both at small and large $B$. The plot shows the coefficient $C$ in the relation $\kappa_{xx}(\mu=0;B)= C\, \alpha^2/b^2 k_B^2T$ as a function of $b/\alpha^2 \sim \omega^{\rm typ}_c \tau_{\rm ee}$. It interpolates between the magnetohydrodynamic regime (\ref{CMHD}) and the large field limit of Eq.~(\ref{ClargeB}).}
\label{fig:kappacrossover}
\end{figure}

\subsection{A.C. response in a magnetic field}

\subsubsection{Cyclotron resonance}
From the expression (\ref{kappafull}) it is easy to infer that all response functions have a pole in the complex frequency plane where the determinant
\bea
\label{zeroofDet}
{\rm Det}[{ R}_0-{R}_1{ G}{R}_1]=0
\eea
vanishes. This defines the cyclotron resonance $\omega^*$ as a function of $B$ and $\rho$. Its trace in the response is illustrated in Fig.~\ref{fig:cyclotron05} where real and imaginary parts of the longitudinal conductivity are plotted as a function of frequency, exhibiting a clear resonance.
\begin{figure}
\includegraphics[width=0.45\textwidth]{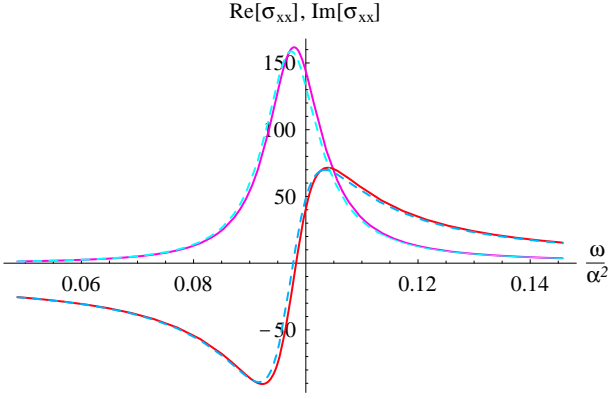}
\caption{Collective cyclotron resonance  evaluated for a small magnetic field $b/\alpha^2=0.5$ and in the quantum critical regime $\mu/T=1$ (light doping). We plot the
real and imaginary part of $\sigma_{xx}(\omega)$ in units of $e^2/\hbar$ as a function of frequency. The red curves were obtained from the full solution of the Boltzmann equation, while the blue curves are the magnetohydrodynamic prediction. The latter relies on the smallness of $b/\alpha^2$, which is seen to be an excellent approximation for the parameter chosen here.
The maximum in the real part occurs at the collective cyclotron frequency $\omega_c=\rho B/(\eP)$. The collisions introduce an intrinsic damping and lead to a broadening of the resonance scaling like $B^2$ at small enough fields.
}
\label{fig:cyclotron05}
\end{figure}

The above condition (\ref{zeroofDet}) is equivalent to
\bea
\omega^*&=&\frac{\rho B}{\varepsilon+P}\\
&& -i\left[g_1(\omega^*,B)-ig_2(\omega^*,B)\right]\frac{B^2}{\eP}\left(I_+^{(1)}-\rho \omega^*\right)^2.\nn
\eea

At small $B$ the solution is given by
\bea
\label{cyclotronres}
\omega^*&\equiv& \omega_c-i\gamma\\
&=&\omega_c^{(0)}-i\gamma^{(0)}\nn\\
&& \quad+\gamma^{(0)}\left[g_1(0,0) I_+^{(1)}\omega_c^{(0)}-\frac{g_2(0,B)}{g_1(0,0)}\right]+{\cal O}(B^4),\nn
\eea
with
\bea
\label{wc0gamma0}
\omega_c^{(0)}=\frac{\rho B}{\eP}\quad; \quad \gamma^{(0)}=\sigma_Q(\mu,0)\frac{B^2}{\eP}\,,
\eea
where $\sigma_Q(\mu,\omega=0)$ was given in (\ref{sigmaQmu}). Note that the expression to order ${\cal O}(B^2)$ is in precise agreement with the predictions from magnetohydrodynamics, cf. Eq.~(\ref{omegac}). However, here we have the additional benefit of obtaining precise information on the dependence of the broadening $\gamma\sim \sigma_Q$ on the chemical potential, as well as on the leading corrections in large fields.

The evolution of the cyclotron resonance with increasing magnetic field is shown in Fig.~\ref{fig:cyclotronmulti} for three different values of small fields. Note that the peak value of the resonance decreases as $1/B^3$, while its width increases as $B^2$.
\begin{figure}
\includegraphics[width=0.45\textwidth]{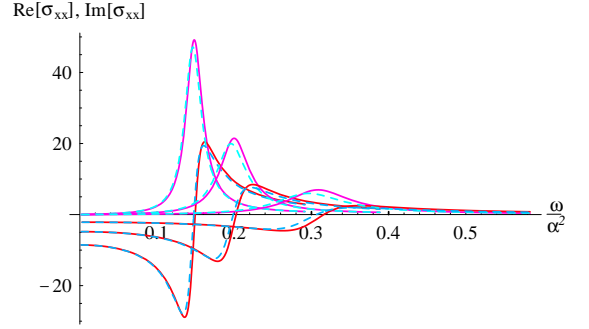}
\caption{Real and imaginary part of $\sigma_{xx}(\omega)$ (in units of $e^2/\hbar$ for $\mu/T=1$ in magnetic fields $b/\alpha^2=0.75, 1, 1.5$ (from left to right). The cyclotron resonance $\omega_c$ is essentially proportional to $B$. The pair of red curves associated to each field value corresponds to the full solution of the Boltzmann equation. The blue curves are the prediction of magnetohydrodynamics which is seen to be an excellent approximation at these small field strengths.}
\label{fig:cyclotronmulti}
\end{figure}

We can also include the effect of weak disorder in the determination of the cyclotron resonance. In the most general case we would have to look for a zero of
\bea
\label{zeroofDet2}
{\rm Det}[{\cal R}_0-{\cal R}_1{\cal G}\overline{{\cal R}}_1]=0
\eea
in the complex frequency plane.

However, similarly as in the calculation of the impurity scattering rate $\tau_{\rm imp}^{-1}$ (\ref{tauimp}), one can check that to lowest order in $\Delta$, only the impurity scattering of the momentum mode enters via the matrix element $M_0^{\rm imp}$, and we can restrict ourselves to the analysis of (\ref{zeroofDet}) with modified $M_0$. Including the latter into the equation for $\omega_*$, one finds that the pole is simplify shifted by $i\tau_{\rm imp}^{-1}$. Such an effect was anticipated by the inclusion of a phenomenological momentum relaxation rate in the relativistic hydrodynamics~\cite{nernst,mhd}, and is put on a rigorous basis here. Of course, the above discussion of disorder effects is meaningful only as long as the impurity scattering rate is the smallest scale in the problem, i.e., for $\Delta\ll b,\alpha^2$.

To avoid confusion, we point out that the collective resonance frequency $\omega_c$ differs from $\omega_c^{\rm typ}$ (\ref{wctyp}), which we had defined as the cyclotron frequency of non-interacting particles at thermal energies. The difference is particularly marked in the critical regime $\mu\ll T$ where $\omega_c$ is significantly smaller than the cyclotron frequency of non-interacting thermal particles. It describes the orbiting motion of the strongly colliding relativistic plasma as a whole, where single particles do not have enough time to complete an orbit, but constantly collide with others, undergoing a collective motion with average frequency $\omega_c\sim \rho$ proportional to the doped density.

\subsubsection{Large fields - beyond hydrodynamics}
Inspecting (\ref{cyclotronres}), an estimate for the field strength where higher order corrections in $B$ become important can be obtained from either of the conditions
\bea
\label{smallB}
\gamma^{(0)}g_1 I_+^{(1)} \sim 1, \quad {\rm or,}\quad \gamma^{(0)}\frac{g_2}{g_1}\sim \omega_c^{(0)}.
\eea
One can check that both are equivalent to $b/\alpha^2\sim 1$, which in turn is equivalent to the physically intuitive condition
\bea
\omega_c^{\rm typ}\tau_{\rm ee}\sim 1,
\eea
both in the quantum critical and the Fermi liquid regime.

In order to characterize the large field regime, we have determined the precise location of the cyclotron resonance by solving numerically for the zero of the determinant (\ref{zeroofDet}) for $\mu/T=1$, relying again on 12 basis functions to calculate $g_{1,2}(\omega,B)$. The result is plotted in Figs.~\ref{fig:Reomegac} and \ref{fig:Imomegac}. We can interpret the increase of $\omega_c$ beyond the hydrodynamic expression as an effect of $\omega_c^{\rm typ}$ becoming of the order of $\tau_{\rm ee}^{-1}$: The larger the field, the further single particles proceed in their unperturbed, non-interacting cyclotron orbits. Since the cyclotron frequency of non-interacting thermal particles is typically higher than $\omega_c$, one should expect that the decrease of the scattering probability leads to an increase of the observed cyclotron resonance frequency, as is seen in Fig.~\ref{fig:Reomegac}.
Given that the scattering is not very efficient in this large field regime, it is also natural that the hydrodynamic approach overestimates the broadening of the resonance due to inelastic collisions, cf. Fig.~\ref{fig:Imomegac}.
\begin{figure}\label{fig:Reomegac}
\includegraphics[width=0.45\textwidth]{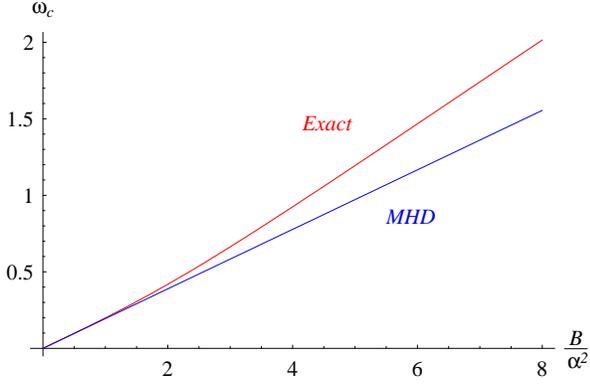}
\caption{Real part of the cyclotron pole $\omega^*\equiv \omega_c-i\gamma$, in units of $k_BT/\hbar\alpha^2$, as a function of magnetic field for fixed chemical potential $\mu/T=1$. The low field part is correctly predicted by relativistic magnetohydrodynamics, while at high fields $\omega_c$ tends to exceed $\omega_c^{\rm MHD}$}.
\end{figure}
\begin{figure}\label{fig:Imomegac}
\includegraphics[width=0.45\textwidth]{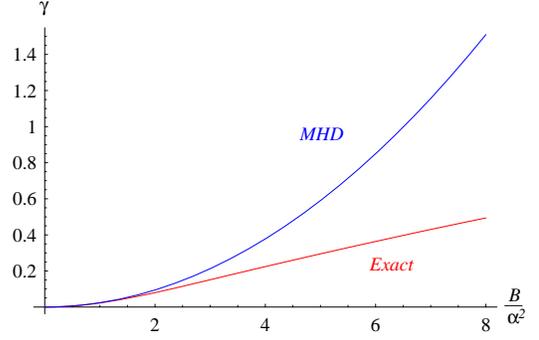}
\caption{Imaginary part of the cyclotron pole $\omega^*\equiv \omega_c-i\gamma$, in units of $k_BT/\hbar\alpha^2$, as a function of magnetic field for fixed chemical potential $\mu/T=1$. The damping at small fields is correctly predicted by relativistic magnetohydrodynamics, while it is overestimated at large values of $b/\alpha^2$.}
\end{figure}

As we pointed out, the magnetohydrodynamic predictions break down at large fields. This is illustrated in Fig.~\ref{fig:cyclotron10} for the cyclotron resonance in the conductivity which is significantly shifted with respect to the hydrodynamic prediction.
\begin{figure}
\includegraphics[width=0.45\textwidth]{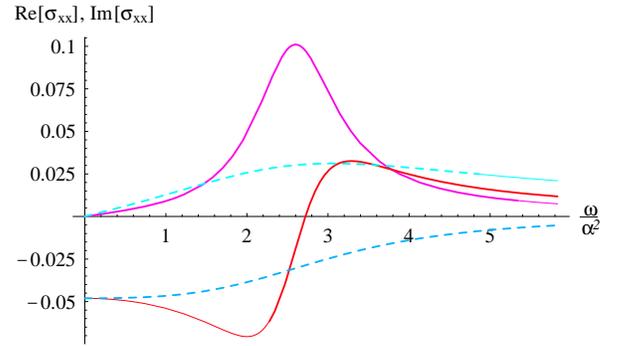}
\caption{Real and imaginary part of $\sigma_{xx}(\omega)$ in units of $e^2/\hbar$ in a large magnetic field $b/\alpha^2=10$, at $\mu/T=1$. The pair of red curves correspond to the full solution of the Boltzmann equation. The blue curves are the prediction of magnetohydrodynamics which fails completely for this large value of $b/\alpha^2\gg 1$.}
\label{fig:cyclotron10}
\end{figure}

The result (\ref{cyclotronres}) for the cyclotron pole is generally valid at small fields for any relativistic fluid, as was shown by the relativistic hydrodynamic analysis in Refs.~\onlinecite{nernst,mhd}. Quite remarkably, this cyclotron resonance also emerged from the solution of an exactly soluble, but {\em strongly} coupled supersymmetric conformal field theory via the AdS-CFT mapping~\cite{AdsCFT}. Furthermore, the deviations from the hydrodynamic prediction at high fields could be found numerically in that case as well, and the trends of $\omega_c(B),\gamma(B)$ were found to be very similar to those in Figs.~\ref{fig:Reomegac},\ref{fig:Imomegac}.
This is very interesting since in the present work we are limited to weak coupling $\alpha\ll 1$, for the Boltzmann approach 
to be quantitatively accurate.

We note that the cyclotron resonance should be readily observable in graphene
at $T$ of the order of room temperature and in moderate magnetic fields corresponding to fractions of a Tesla, as discussed in Ref.~\onlinecite{mhd}

\subsection{Recovery of Kohn's theorem}
The above collective cyclotron  effects are most pronounced in the relativistic, quantum critical regime $|\mu|\lesssim T$. As one leaves the latter, the cyclotron resonance at a given magnetic field $B$ becomes sharper and sharper as characterized by the ratio $\gamma/\omega_c$ which equals the number of applied flux quanta per doped carrier, multiplied by $\sigma_Q(\mu)$ (as measured in units of $e^2/\hbar$). As we have seen above, $\sigma_Q(\mu)$ decays as $(T/\mu)^{2}$ at large $\mu$. The resulting sharpening of the resonance is clearly seen in Fig.~\ref{fig:cyclotron_mu5} which shows $\sigma_{xx}(\omega)$ evaluated for $\mu/T=5$.
\begin{figure}
\includegraphics[width=0.45\textwidth]{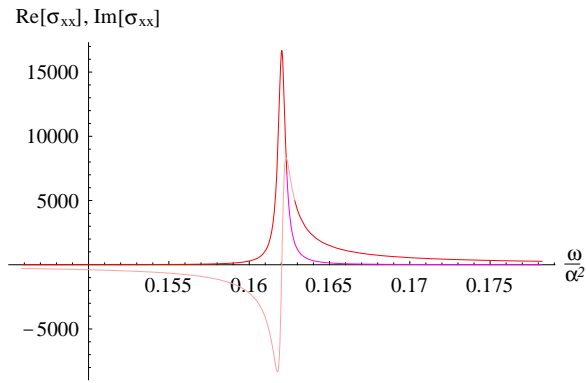}
\caption{Sharp cyclotron resonance in the conductivity (in units of $e^2/\hbar$) for $\mu/T=5$ and a moderate magnetic field $b/\alpha^2=1$. Since in the Fermi liquid regime the presence of antiparticles (holes) can be neglected, and the relevant part of the single particle dispersion is hard to distinguish from a parabolic dispersion, Kohn's theorem applies approximately, the cyclotron resonance becoming sharper and sharper at asymptotically large $\mu/T$.}
\label{fig:cyclotron_mu5}
\end{figure}

It is interesting to note that as $\mu/T$ increases, the resonance approaches the value
\bea
\omega_c^{(0)}=\frac{\rho B}{\eP}\to \frac{eB}{\mu/v_F^2}=\frac{eB}{\hbar k_F/v_F},
\eea
which one recognizes as the semiclassical cyclotron frequency expected for a circular Fermi surface at wavevector $k_F$ and Fermi velocity $v_F$. Both observations indicate that one recovers the familiar Fermi liquid characteristics at large doping.

The intrinsic broadening of the cyclotron resonance due to collisions is an interesting effect pertaining mostly to the quantum critical regime $|\mu|\lesssim T$.
As noted above, in the Fermi liquid regime the width $\gamma$ of the cyclotron resonance tends to zero. This can be understood as reflecting the approach of a regime where Kohn's theorem  should apply asymptotically: One species of quasiparticles is entirely frozen out in this regime, and the deviation of the linear band structure from a parabolic dispersion becomes increasingly negligible. These are the two conditions under which Kohn's theorem is valid: The latter  asserts that in a system with only one parabolic band there is a single sharp resonance peak at a well defined cyclotron frequency, irrespective of the presence of electron-electron interactions.

\subsection{Range of validity of relativistic magnetohydrodynamics}
Using Eqs.~(\ref{01Equation_1},\ref{kappafull}) one can obtain explicit expressions for the frequency dependent response in pure systems in a magnetic field. Note the remarkable fact that the three pairs of longitudinal and transverse response coefficients are strongly constrained: Apart from thermodynamic data such as $(\eP),s,T,\rho$ and $\mu$ the formulae contain only two
independent matrix elements $g_{1,2}(\omega,B,\mu)$.
To leading order in magnetic field one can even neglect the dependence on $g_2$, and all the response functions are interrelated, with one single parameter left, corresponding directly to $\sigma_Q$ in the hydrodynamic formulation.

The full expressions for the response functions can easily be worked out analytically, but the expressions are relatively lengthy and will not be displayed here.
However, it is interesting to use these exact results to determine the extent to which the magnetohydrodynamic formulae given in Sec.~\ref{MHD} are valid. As we already know, at large fields corrections set in, and similar corrections are to be expected at higher frequencies of the order of the inelastic scattering rate.
We are now in the position to characterize the corrections precisely.
A detailed analysis of the response shows that the following statement holds: For small frequencies ($\omega/\alpha^2\ll 1$) and small fields ($b/\alpha^2\ll 1$), and a fixed ratio $\omega/b$, the exact a.c. conductivity satisfies the asymptotic equality
\bea
\sigma_{xx}(\omega,B)= \sigma^{\rm MHD}_{xx}(\omega,B)+ {\cal O}(b/\alpha^2,\omega/\alpha^2).
\eea
An analogous relation holds for all other response functions. Hereby the expressions (\ref{wc0gamma0}), with $\sigma_Q$ from (\ref{sigmaQmu}), have to be used in the magnetohydrodynamic response functions given in Sec.~\ref{MHD}.

It is interesting that a very similar result was obtained from the exact solution of the strongly coupled conformal field theory studied in Ref.~\onlinecite{AdsCFT}, showing that the validity of this statement is not restricted to weak coupling.

\section{Conclusion}
We have presented a Boltzmann approach to describe transport in liquids of interacting Dirac fermions with and without magnetic fields. We have established that as long as the inelastic scattering rate is the largest scattering rate in the problem, the relativistic hydrodynamic formalism captures the frequency-dependent response very well. Further, we have obtained an exact expression for the single transport coefficient $\sigma_Q$ that is left undetermined by hydrodynamics, and showed that it decays as a power law as one leaves the quantum critical relativistic regime. At large doping the electron system was shown to recover all the signatures of a Fermi liquid, such as Mott's law and the Wiedemann Franz relation.
At the same time the collective cyclotron resonance, a remarkable feature of quantum criticality, turns gradually into a sharp resonance centered at the semiclassical cyclotron frequency as one dopes the system further.

Finally an analysis of the large field behavior yielded similar qualitative deviations of the cyclotron pole from the corresponding hydrodynamic prediction as was found in the exact solution of a strongly coupled conformal field theory.

\acknowledgments We thank Joe Bhaseen, S. Hartnoll, B. Rosenow, J.~Schmalian and B. Shklovskii for discussions. This research was supported by the Swiss National
Fund for Scientific Research under grant PA002-113151 (MM); by the Deutsche
Forschungsgemeinschaft under grant FR 2627/1-1 (LF), and by the NSF under
grant DMR-0537077 (SS).

\appendix
\section{Relativistic hydrodynamics}
\label{app:HD}
In this appendix we discuss the hydrodynamic and constitutive equations which are used to obtain the hydrodynamic response functions given in the main part of the text.

In covariant notation, the conservation laws for a relativistic fluid read
\begin{eqnarray}
\label{current}
\p_\alpha {J}^{\alpha} &=& 0,\\
\label{Tmunu}
\p_\beta {T}^{\alpha\beta}&=& F^{\alpha\gamma}{J}_\gamma,
\end{eqnarray}
where the energy-momentum tensor and current vector of the fluid are given by
\begin{eqnarray}
T^{\alpha\beta} &=& (\vep+P) u^{\alpha} u^\beta + P g^{\alpha\beta} +
\tau^{\alpha\beta},
\label{e0} \\
J^\alpha &=& \rho u^\alpha + \nu^\alpha,
\end{eqnarray}
where $\vep$ is the energy density, $P$ is the pressure, $\rho$ the charge density, $g^{\mu\nu}={\rm diag}(-1,1,1)$ the Lorentz metric and $F^{\mu\nu}$ is the electromagnetic field tensor. Note that the role of the speed of light is taken by the Fermi velocity $v_F$. The velocity field $u^\mu$ (in units of $v_F$) is determined in such a way that there is no {\em energy} flow in the local rest frame where $u^\mu=(1,\mathbf{0})$. Notice that due to the presence of heat flows this does not coincide in general with the velocity defined by the average  {\em charge} current.

The additional terms in (\ref{current},\ref{Tmunu}) are dissipative contributions:  the Reynolds tensor $\tau^{\mu\nu}$
accounts for viscous forces, which turn out to be irrelevant for small wavevector response. The vector $\nu^\mu$ is proportional to the heat current.
To obtain a closed set of equations, one has to express the heat current in terms of local quantities. In a relativistic system, its form is strongly constrained by covariance and the requirement that the entropy of the liquid always increases. The divergence of the entropy current follows from the equations of motion as:
\bea
\label{entropyprod}
\d_\alpha \left(su^\alpha - \frac\mu T \nu^\alpha\right) &=&
   -\nu^\alpha \left[\d_\alpha \left(\frac\mu T\right) - \frac{1}{T} F_{\alpha\gamma} u^\gamma\right]\nn\\
&&\quad - \frac{\tau^{\alpha\gamma}} T \d_\alpha u_\gamma\,,
\eea
where we have used the thermodynamic identity $\eP-\mu\rho =s T$ for the entropy density, $s$.
The requirement that the left hand side be positive, and the assumption that the heat current should be linear in the gradients of $T$ and $\mu$, as well as in the electromagnetic fields (i.e., the gradients of the scalar and vector potential), imposes that
\bea
\nu^\alpha&=-\sigma_Q(g^{\alpha\lambda}+u^\alpha
u^\lambda)\left[T\partial_\lambda\left(\frac{\mu}{T}\right)-F_{\lambda\gamma}u^\gamma\right].
\label{nu-constit}
\eea
A similar relation holds for $\tau^{\alpha\gamma}$. This leaves us with a single undetermined transport coefficient $\sigma_Q>0$ with units of a conductivity. Note that in the relativistic case the heat current is not only proportional to the thermal gradient, but also to the acceleration of the fluid element (second term). Note that the assumption that $\nu^\alpha$ is linear in $F^{\mu\nu}$ restricts the above argument to small fields $B$.

\section{Matrix elements of terms in the Boltzmann equation}
\label{app:Melements}
\subsection{Matrix elements in a general basis}
In this appendix we give explicit expressions for the matrix elements appearing in the Boltzmann equation after a projection onto a specific basis of modes. The collision integral describing electron-electron scattering as well as impurity scattering is given by
\begin{widetext}
\begin{eqnarray}
&&{\cal I}_{\rm coll}[\lambda,k,t\,|\{f\}]=2\pi \alpha^2
\int \frac{d^{2}k_{1}}{(2\pi )^{2}}\frac{d^{2}q}{(2\pi )^{2}}\Biggl\{
\\
&&\quad\delta (k-k_{1}-|\mathbf{k}+\mathbf{q}|+|\mathbf{k}_{1}-\mathbf{q}|)R_{1}(
\mathbf{k},\mathbf{k}_{1},\mathbf{q})\Bigl\{f_{\lambda }(\mathbf{k}
,t)f_{-\lambda }(\mathbf{k}_{1},t)[1-f_{\lambda }(\mathbf{k}+\mathbf{q},t)]
\notag \\
&&\quad~~~~~~\times \lbrack 1-f_{-\lambda }(\mathbf{k}_{1}-\mathbf{q}
,t)]-[1-f_{\lambda }(\mathbf{k},t)][1-f_{-\lambda }(\mathbf{k}
_{1},t)]f_{\lambda }(\mathbf{k}+\mathbf{q},t)f_{-\lambda }(\mathbf{k}_{1}-
\mathbf{q},t)\Bigr\}  \notag \\
&&\quad\delta (k+k_{1}-|\mathbf{k}+\mathbf{q}|-|\mathbf{k}_{1}-\mathbf{q}|)R_{2}(
\mathbf{k},\mathbf{k}_{1},\mathbf{q})\Bigl\{f_{\lambda }(\mathbf{k}
,t)f_{\lambda }(\mathbf{k}_{1},t)[1-f_{\lambda }(\mathbf{k}+\mathbf{q},t)] \notag \\
&&\quad~~~~~~\times \lbrack 1-f_{\lambda }(\mathbf{k}_{1}-\mathbf{q}
,t)]-[1-f_{\lambda }(\mathbf{k},t)][1-f_{\lambda }(\mathbf{k}
_{1},t)]f_{\lambda }(\mathbf{k}+\mathbf{q},t)f_{\lambda }(\mathbf{k}_{1}-
\mathbf{q},t)\Bigr\}\Biggr\}   \notag  \\
&&\quad
+2\pi \int \frac{d^2 k_1}{(2\pi)^2} \delta(k-k_1) |U_{\lambda \lambda}|^2 \Big[ f_\lambda (\vk,t)(1-f_\lambda(\vko,t))-(1-f_\lambda(\vk,t))f_\lambda (\vko,t) \Big]\,, \notag
\end{eqnarray}
where
\begin{eqnarray}
R_{1}(\mathbf{k},\mathbf{k}_{1},\mathbf{q}) &=&\frac{4}{\alpha^2}\left( \bigl|T_{+--+}(
\mathbf{k},\mathbf{k}_{1},\mathbf{q})-T_{+-+-}(\mathbf{k},\mathbf{k}_{1},-
\mathbf{k}-\mathbf{q}+\mathbf{k}_{1})\bigr|^{2}\right.  \notag \\
&&\left. ~+(N-1)\bigl|T_{+--+}(\mathbf{k},\mathbf{k}_{1},\mathbf{q})\bigr|
^{2}+(N-1)\bigl|T_{+-+-}(\mathbf{k},\mathbf{k}_{1},-\mathbf{k}-\mathbf{q}+
\mathbf{k}_{1})\bigr|^{2}\right)\,,  \notag \\
R_{2}(\mathbf{k},\mathbf{k}_{1},\mathbf{q}) &=&\frac{4}{\alpha^2}\left( \frac{1}{2}\bigl|
T_{++++}(\mathbf{k},\mathbf{k}_{1},\mathbf{q})\ -T_{++++}(\mathbf{k},\mathbf{
k}_{1},\mathbf{k}_{1}-\mathbf{k}-\mathbf{q})\ \bigr|^{2}\right.  \notag \\
&~&~~~~\left. ~~~~+(N-1)\bigl|T_{++++}(\mathbf{k},\mathbf{k}_{1},\mathbf{q})
\bigr|^{2}\right)\,,  \label{deft12}
\end{eqnarray}
and the disorder potential $U_{\lambda \lambda}$ was introduced in Eq.~\eqref{dispot}.
The inelastic scattering rate is proportional to $\alpha^2$, and the matrix entries of the electron-electron scattering matrix ${\cal M}^{\rm{Cb}}$ read explicitly
\bea
\label{MCb}
{\cal M}_{mn}^{\rm{Cb}}&=& 2\pi \alpha^2 \sum_\lambda  \int \frac{d^{2}k}{(2\pi )^{2}}\frac{
d^{2}k_{1}}{(2\pi )^{2}}\frac{d^{2}q}{(2\pi )^{2}}\delta (k+k_{1}-|\mathbf{k}+\mathbf{q}|-|\mathbf{k}_{1}-\mathbf{q}|)   \phi_m (\lambda,k) \\
&&\Biggl\{R_{1}(\mathbf{k},\mathbf{q}-\mathbf{k}_{1},\mathbf{q})
(1-f^0_{\lambda k})(1-f^0_{-\lambda |\vko-\vq|})f^0_{-\lambda k_1} f^0_{\lambda |\vk+\vq|}\notag \\
&&~\times {\bf{e}}_{\bf{k}}  \cdot \left[ {\bf{e}}_{\bf{k}}\phi_n(\lambda,k)+{\bf{e}}_{\bf{k}_1}
\phi_n(-\lambda,k_{1})-{\bf{e}}_{\bf{k}+\bf{q}}
\phi_n(\lambda,|\mathbf{k}+\mathbf{q}|)-{\bf{e}}_{\bf{k}_1-\bf{q}} \phi_n(-\lambda,|\mathbf{k}_{1}-\mathbf{q}|)\right]   \notag\\
&&+R_{2}(\mathbf{k},\mathbf{k}_{1},\mathbf{q}) (1-f^0_{\lambda k}) (1-f^0_{\lambda k_1})f^0_{\lambda |\vk+\vq|} f^0_{\lambda |\vko-\vq|} \notag \\
&&~\times {\bf{e}}_{\bf{k}}  \cdot\left[ {\bf{e}}_{\bf{k}} \phi_n(\lambda,k)+{\bf{e}}_{\bf{k}_1}
\phi_n(\lambda,k_{1})-{\bf{e}}_{\bf{k}+\bf{q}}\phi_n(\lambda,|
\mathbf{k}+\mathbf{q}|)-{\bf{e}}_{\bf{k}_1-\bf{q}} \phi_n(\lambda,|\mathbf{k}_{1}-\mathbf{q}|)\right]\Biggr\}\; ,
\notag
\eea
\end{widetext}
where
\begin{eqnarray}
f^0_{\lambda k}:=\frac{1}{e^{\lambda k-\mu}+1}\; .
\end{eqnarray}

The entries in the impurity scattering matrix $\mathcal{M}^{\rm{imp}}$ are given by
\begin{eqnarray}
\mathcal{M}_{mn}^{\rm{imp}}=2\pi \Delta \sum_\lambda   \int \frac{d^{2}k}{(2\pi )^{2}}\frac{d^{2}k_{1}}{(2\pi )^{2}}\frac{\delta (k-k_1)}{|{\bf{k}}-{\bf{k}}_1|^2}\quad\\
\quad\times f^0_{\lambda k} (1-f^0_{\lambda  k_1}) \phi_m(\lambda,k) \left [\phi_n (\lambda,k) - {\bf{e}}_{\bf{k}} \cdot{\bf{e}}_{\bf{k}_1} \phi_n(\lambda,k_1)  \right]\,,\nn
\end{eqnarray}
where the strength of disorder is measured by the dimensionless parameter
\begin{eqnarray}
\Delta=\pi^2 \left ( \frac{Z e^2}{k_BT\epsilon_r}\right)^2 n_{{\rm {imp}}}\,.
\end{eqnarray}
The matrix for the time derivative reads
\begin{eqnarray}
\mathcal{M}_{mn}^{i\omega}=i\omega \sum_\lambda \int \frac{d^{2}k}{(2\pi )^{2}} f^0_{\lambda k}(1-f^0_{\lambda k})   \phi_m (\lambda,k)\phi_n (\lambda,k) \;.\nonumber \\
\end{eqnarray}
Finally, the deflection of currents by the magnetic field is described by a matrix with entries
\begin{eqnarray}
\mathcal{B}_{mn}=b\sum_\lambda \int \frac{d^2k}{(2\pi)^2} f^0_{\lambda k}(1-f^0_{\lambda k})\frac{\lambda}{k}\phi_m (\lambda,k) \phi_n (\lambda,k)\;,\nonumber \\
\end{eqnarray}
where the dimensionless parameter measuring the field strength is given by
\begin{eqnarray}
b=\frac{eBv_F^2}{(k_BT)^2}\;.
\end{eqnarray}

The projection of the driving terms due to an electric field or a temperature gradient onto the basis functions yields the vectors
\begin{eqnarray}
\label{eq:force}
\mathcal{F}^E_m &=&  \sum_\lambda \lambda \int \frac{d^2 k }{(2\pi)^2} f^0_{\lambda k}(1-f^0_{\lambda k})  \phi_m (\lambda,k),  \\
\mathcal{F}^T_m &=& -\sum_\lambda \int \frac{d^2 k }{(2\pi)^2} \left ( k - \lambda \mu\right) f^0_{\lambda k}(1-f^0_{\lambda k}) \phi_m (\lambda,k)\;.\nonumber
\end{eqnarray}

\subsection{Two mode approximation}
Here, we evaluate the matrix elements with respect to the two main modes, $\phi_0,\phi_1$, of the specific basis (\ref{modes}-\ref{orthogonality}).
In the sector spanned by these two modes, the electron-electron collision operator takes the form
\bea
{\cal M}^{\rm Cb}=\left(\begin{array}{cc} 0 & 0\\ 0 & {\bf M}^{\rm Cb}_{11}({\mu}) \end{array}\right),
\eea
where
\bea
\label{W}
&&\mathbf{M}^{\rm Cb}_{11}({\mu})=
\alpha^2\sum_\lambda \frac{2\pi }{4} \int \frac{d^{2}k}{(2\pi )^{2}}\frac{d^{2}k_{1}}{(2\pi )^{2}}\frac{d^{2}q}{(2\pi )^{2}}\Biggl\{  \notag \\
&&\quad\delta (k-k_{1}-|\mathbf{k}+\mathbf{q}|+|\mathbf{k}_{1}-\mathbf{q}|){R}_{1}(\mathbf{k},\mathbf{k}_{1},\mathbf{q})  \notag \\
&&\quad~\times (1-f^0_{\lambda k})(1-f^0_{-\lambda k_1})f^0_{\lambda |\mathbf{k}+\mathbf{q}|} f^0_{-\lambda|\mathbf{k}_{1}-\mathbf{q}|} \nn\\
&&\quad~\times\left[\frac{\mathbf{k}}{k}-\frac{\mathbf{k}_{1}}{k_{1}}-\frac{(\mathbf{k}+\mathbf{q})}{|\mathbf{k}+\mathbf{q}|}+\frac{(\mathbf{k}_{1}-\mathbf{q})}{|\mathbf{k}_{1}-\mathbf{q}|}\right] ^{2}  \notag\\
&&\quad+\delta (k+k_{1}-|\mathbf{k}+\mathbf{q}|-|\mathbf{k}_{1}-\mathbf{q}|){R}_{2}(\mathbf{k},\mathbf{k}_{1},\mathbf{q})\nn \\
&&\quad~\times (1-f^0_{\lambda k})(1-f^0_{\lambda k_1})f^0_{\lambda |\mathbf{k}+\mathbf{q}|}f^0_{\lambda|\mathbf{k}_{1}-\mathbf{q}|} \nn\\
&&\quad~\times\left[ \frac{\mathbf{k}}{k}+\frac{\mathbf{k}_{1}}{k_{1}}-\frac{(\mathbf{k}+\mathbf{q})}{|\mathbf{k}+\mathbf{q}|}-\frac{(\mathbf{k}_{1}-\mathbf{q})}{|\mathbf{k}_{1}-\mathbf{q}|}\right] ^{2}\Biggr\} \;.
\end{eqnarray}

The matrix elements of all other operators can be expressed with the help of the functions
\begin{eqnarray}
&&I^{(n)}_{s=\pm}({\mu}) = \frac{N}{2}\sum_\lambda \int \frac{d k}{2\pi}\, \left(\delta_{s,+}+\lambda\delta_{s,-}\right) k^n f^0_{\lambda k}(1-f^0_{\lambda k}) \quad\nn\\
&&\quad= \frac{N}{2}\left[\frac{\delta_{n,0}\delta_{s,+}}{2\pi}+n \sum_\lambda \int \frac{d k}{2\pi}\, \left(\delta_{s,+}+\lambda\delta_{s,-}\right) k^n f^0_{\lambda k}\right],\nn
\eea
where a partial integration was used to obtain the second line.
One easily verifies the explicit relations
\bea
I^{(0)}_+ &=& \frac{N}{2}\frac{1}{2\pi},\\
I^{(0)}_- &=& \frac{N}{2}\frac{\tanh({\mu}/2)}{2\pi},\\
I^{(1)}_+  &=& \frac{N}{2}\frac{1}{\pi}\ln[2 \cosh({\mu}/2)],\\
I^{(1)}_-  &=& \frac{N}{2}\frac{{\mu}}{2\pi},\\
I^{(2)}_+ &=& \rho^+ +\rho^-,\\
I^{(2)}_-  &=&  \rho^+ -\rho^-=\rho,\\
I^{(3)}_+ &=& \eP =\frac{3}{2}\varepsilon,
\eea
where $\rho^\pm$ are the number densities of particles and holes, respectively, and $\rho$ is the charge density in units of $e$. The last relation follows since in a two-dimensional relativistic liquid $P=\varepsilon/2$. This is a consequence of the energy momentum tensor being traceless, a relation that can easily be checked explicitly for free Dirac fermions.

In order to analyze the Fermi-liquid regime it is convenient to have the asymptotic form of $I^{(n)}_\lambda$ for large ${\mu}$ at hand. From a standard Sommerfeld expansion one finds
\begin{eqnarray}\label{eq:Fermi}
I^{(0)}_\lambda ({\mu}) &\approx&  \frac{N}{2}\frac{1}{2\pi}\,, \nonumber \\
I^{(1)}_\lambda ({\mu}) &\approx&  \frac{N}{2}\frac{{\mu}}{2\pi}\,, \nonumber \\ I^{(2)}_\lambda ({\mu}) &\approx& \frac{N}{2} \left[ \frac{{\mu}^2}{2\pi}+\frac{\pi}{6}\right]\,,\nonumber\\
I^{(3)}_\lambda ({\mu}) &\approx& \frac{N}{2}\left[ \frac{{\mu}^3}{2\pi}+\frac{\pi}{2}{\mu}\right]\,,
\end{eqnarray}
up to corrections of order ${\cal O}(\exp(-\mu))$.

In the two mode approximation, the scattering from Coulomb impurities is described by
\begin{eqnarray}
{\cal M}^\textrm{imp}= \frac{2}{N}\Delta \left( \begin{array} {cc} I_+^{(2)}({\mu})  & I_-^{(1)}({\mu}) \\ I_-^{(1)} ({\mu})  & I_+^{(0)} ({\mu} ) \end{array}\right),
\end{eqnarray}
and the matrix for the time derivative takes the form
\begin{eqnarray}
{\cal M}^{i\omega}=
-i\omega\frac{2}{N} \left ( \begin{array}{cc} \eP & \rho \\ \rho &  I_+^{(1)}({\mu})\end{array}\right).
\end{eqnarray}
Finally, the matrix describing the deflection by the magnetic field has the two mode representation
\bea
{\cal B}= b\frac{2}{N} \left ( \begin{array}{cc} \rho & I^{(1)}_+({\mu}) \\   I_+^{(1)}({\mu}) & I^{(0)}_-({\mu})\end{array}\right).
\eea

Due to the choice (\ref{orthogonality}) the driving fields only have components along $\phi_0$ and $\phi_1$ given by
\begin{eqnarray}
\vec{{\cal F}}^{E}= \frac{2}{N}\left ( \begin{array}{c} \rho \\ I_+^{(1)}({\mu}) \end{array}\right)\,,
\end{eqnarray}
and,
\begin{eqnarray}
\vec{{\cal F}}^{T} &=& \frac{2}{N} \left(\begin{array}{c}  \eP-{\mu} \rho \\ \rho-{\mu} I_+^{(1)}({\mu}) \end{array}\right) \nn\\
&=&\frac{2}{N} \left(\begin{array}{c}  s \\ \rho-\frac{N}{2}\frac{{\mu}}{\pi} \ln[2\cosh({\mu}/2)] \end{array}\right)\,.
\end{eqnarray}

\section{Inelastic scattering rate in the Fermi liquid regime}
\label{app:inel}
In this appendix we analyze the behavior of the matrix element ${\bf M}_{11}^{\rm Cb}$ in the limit of large chemical potential $\mu\gg T$, in order to estimate the quantity $g_1^{-1}$ which determines the inelastic scattering rate.

The phase space for the outer ${\bf k}$-integral in (\ref{W}) scales like $\mu T$. Further, given a momentum transfer ${\bf q}$, the integral over ${\bf k}_1$ over the energy conserving $\delta$-function scales like $T\mu/|\mathbf{q}|$, where the factor $1/|\mathbf{q}|$ is due to the derivative of the argument of the $\delta$-function. The factors $R_{1,2}$ scale like $\alpha^2/|\mathbf{q}|^2$ if the Coulomb interactions are not screened, and thus, for small $\mathbf{q}$ the integral over the momentum transfer behaves like $\int d^2q q^{-3}$. The infrared divergence is cut off by the square of the difference of distribution functions $g_\lambda(k)$ which provides an extra factor of $q^2$ at small $q$. The cut off scale is set by the typical range over which $g(k+q)$ varies, which is $q\sim T$. Consequently the integral is dominated by $q\sim T$ which contributes a phase space factor $T^2$.
In the final estimate of $g_1^{-1}(\mu)$ we need to take into account that the relevant mode for electrical conductivity, $\phi_1$, has a strong overlap with the zero mode $\phi_0$, when $k$ is restricted to the thermally relevant vicinity of the Fermi level $\mu$. Since $\phi_0$ is a zero mode of the integral, only the part of $\phi_1$ orthogonal to $\phi_0$ contributes, and this provides an extra factor of $(T/\mu)^2$.
Putting all these factors together and multiplying by a normalizing factor $1/T^3$, we find that unscreened electron-electron interactions are dominated by small momentum transfer of order $q\sim T$, leading to a scattering parameter
\bea
g_1^{-1}(\mu)\sim \frac{\alpha^2}{T^3}\left[ \mu T\,\frac{T\mu}{q} \frac{q^2}{|\mathbf{q}|^2}\frac{T^2}{\mu^2}\right]_{q\sim T}= {\cal O}(1),
\eea
which tends to a constant.

The inelastic scattering rate can be estimated from a similar expression, where the first ${\mathbf k}$-integral and the normalization factor are dropped. Looking at the relaxation of modes $g(k)$ that have a variation of order $\mathcal{O}(1)$ over an interval of order $\mathcal{O}(T)$ around $k=\mu$, there is no suppression factor from the square of differences in $g_\lambda(k)$, and we obtain the estimate
\bea
\tau^{-1}_{\rm ee}\sim \alpha^2\left[\frac{T\mu}{q}\frac{q^2}{|\mathbf{q}|^2}\right]_{q\sim T}\sim \alpha^2\mu.
\eea
This is indeed consistent with the expression (\ref{tauee}), upon using the estimate (\ref{g_1estimate}).

If we include screening of the interactions, the scattering is eventually dominated by $q\sim \mu$, while the phase space integral of the $\mathbf{k}_1$-integral contributes $qT$ instead of $q^2$. This leads to
\bea
g_{1,\rm sc}^{-1}(\mu)\sim \frac{\alpha^2}{T^3}\left[ \mu T\,\frac{T\mu}{q} \frac{q T}{|\mathbf{q}|^2}\frac{T^2}{\mu^2}\right]_{q\sim \mu}= {\cal O}\left(\alpha^2\frac{T^2}{\mu^2}\right),
\eea
and the analogous estimate for $\tau_{\rm ee}$ yields
the familiar Fermi liquid behavior
\bea
\tau^{-1}_{\rm ee,sc}\sim \alpha^2\left[\frac{T\mu}{q}\frac{q T}{|\mathbf{q}|^2}\right]_{q\sim \mu}\sim \alpha^2\frac{T^2}{\mu}\,.
\eea


\begin{thebibliography}{99}
\bibitem{footnote1} We have checked that using up to 16 basis functions affected the numerical values by less than a percent, indidcating that good numerical accuracy was achieved with 12 basis functions. The latter were chosen as $\{k^n,\lambda k^n\}$, multiplied by a cut-off factor $\exp[-k/2]$ for powers $n\geq 3$.
\end{thebibliography}
\end{document}